\documentclass[12pt]{article}
\usepackage[nosort]{cite}
\usepackage[usenames, dvipsnames]{xcolor}
\usepackage{graphicx}
\usepackage{multicol}
\usepackage{amsfonts}
\usepackage{amssymb}
\usepackage{amsmath}
\usepackage{heck}
\usepackage{braket}
\usepackage{afterpage}
\usepackage{setspace}
\usepackage{verbatim}
\usepackage{booktabs}
\usepackage[para]{threeparttable}
\usepackage{color}
\usepackage[normalem]{ulem}
\usepackage{longtable}
\usepackage{float}
\usepackage{subcaption}
\usepackage{epsfig}
\usepackage{bm}
\usepackage{enumerate}
\usepackage{epstopdf}
\usepackage[enableskew, vcentermath]{youngtab}
\usepackage{adjustbox}
\usepackage{multirow}
\usepackage{tikz}
\usepackage[margin=1in]{geometry}
\usepackage{titletoc}
\usepackage{hyperref}
\usepackage[percent]{overpic}
\usepackage{mathtools}
\usepackage{tikz-cd}
\providecommand{\U}[1]{\protect\rule{.1in}{.1in}}
\pdfoutput=1
\newsavebox{\mysavebox}

\hypersetup{colorlinks,citecolor=black,filecolor=black,linkcolor=black,urlcolor=black}
\usetikzlibrary{decorations.markings}

\numberwithin{equation}{section}

\usetikzlibrary{chains}
\allowdisplaybreaks
\tikzset{node distance=2em, ch/.style={circle,draw,on chain,inner sep=2pt},chj/.style={ch,join},every path/.style={shorten >=4pt,shorten <=4pt},line width=1pt,baseline=-1ex}

\newcommand{\ba}{\begin{eqnarray}}
\newcommand{\ea}{\end{eqnarray}}

\newcommand{\be}{\begin{equation}}
\newcommand{\ee}{\end{equation}}

\tikzstyle{startstop} = [rectangle, rounded corners, minimum width=3cm, minimum height=1cm,text centered, draw=black, fill=blue!10]
\tikzstyle{startstop} = [rectangle, rounded corners, minimum width=3cm, minimum height=1cm,text centered, draw=black, fill=blue!10]
\tikzstyle{io} = [trapezium, trapezium left angle=70, trapezium right angle=110, minimum width=3cm, minimum height=1cm, text centered, draw=black, fill=blue!30]
\tikzstyle{process} = [rectangle, minimum width=3cm, minimum height=1cm, text centered, draw=black, fill=orange!30]
\tikzstyle{decision} = [diamond, minimum width=3cm, minimum height=1cm, text centered, draw=black, fill=green!30]
\tikzstyle{arrow} = [thick,->,>=stealth]
\tikzset{->-/.style={decoration={
  markings,
  mark=at position #1 with {\arrow[scale=2.4]{>}}},postaction={decorate}}}
\makeatletter \@addtoreset{equation}{section} \makeatother

\begin{document}

\date{August 2022}

\title{Super-Spin Chains for 6D SCFTs}

\institution{PENN}{\centerline{${}^{1}$Department of Physics and Astronomy, University of Pennsylvania, Philadelphia, PA 19104, USA}}
\institution{DESY}{\centerline{${}^{2}$Deutsches Elektronen-Synchrotron DESY, Notkestr.~85, 22607 Hamburg, Germany}}
\authors{Florent Baume\worksat{\PENN}\footnote{e-mail: {\tt fbaume@sas.upenn.edu}}, Jonathan J.\ Heckman\worksat{\PENN}\footnote{e-mail: {\tt jheckman@sas.upenn.edu}}, and Craig Lawrie\worksat{\DESY}\footnote{{e-mail: {\tt craig.lawrie1729@gmail.com}}}}
\abstract{
	Nearly all 6D superconformal field theories (SCFTs) have a partial tensor branch description in terms of a
generalized quiver gauge theory consisting of a long
	one-dimensional spine of quiver nodes with links given by conformal
	matter; a strongly coupled generalization of a bifundamental
	hypermultiplet. For theories obtained from M5-branes probing an ADE
	singularity, this was recently leveraged to extract a protected large
	R-charge subsector of operators, with operator mixing controlled at
	leading order in an inverse large R-charge expansion by an integrable spin $s$
	Heisenberg spin chain, where $s$ is determined by the
	$\mathfrak{su}(2)_{R}$ R-symmetry representation of the conformal
	matter operator. In this work, we show that this same structure extends to the
	full superconformal algebra $\mathfrak{osp}(6,2| 1)$. In
	particular, we determine the corresponding Bethe ansatz equations which
	govern this super-spin chain, as well as distinguished subsectors which
	close under operator mixing. Similar considerations extend to 6D
	little string theories (LSTs) and 4D $\mathcal{N} = 2$ SCFTs with the
	same generalized quiver structures.
}

\begin{flushright}
  {\small \texttt{\hfill DESY-22-132}}
\end{flushright}

\maketitle

\setcounter{tocdepth}{2}

\tableofcontents

\enlargethispage{\baselineskip}

\newpage

\section{Introduction}\label{sec:INTRO}

The study of 6D superconformal field theories (SCFTs) is
undergoing a renaissance. While allowed on general representation theory
grounds \cite{Nahm:1977tg}, the first examples of interacting 6D SCFTs only
appeared in the mid-1990s due to an improved understanding of supersymmetric
string and gauge theory dynamics \cite{Witten:1995zh,
Strominger:1995ac,Seiberg:1996qx}, and the first known examples were recognized as
interacting fixed points in \cite{Seiberg:1996qx}. The recent renewed interest in the subject
has been spurred on by the development of new systematic techniques for the
construction and study of such theories.\footnote{See e.g., \cite{Heckman:2013pva,DelZotto:2014hpa, Heckman:2014qba, Intriligator:2014eaa,
Ohmori:2014pca, Ohmori:2014kda, DelZotto:2014fia, Heckman:2015bfa,
Bhardwaj:2015xxa, Tachikawa:2015wka, DelZotto:2015isa, Bhardwaj:2015oru,
Cordova:2018cvg, Bhardwaj:2018jgp, Heckman:2018pqx,
Bhardwaj:2019hhd,Bergman:2020bvi,Baume:2020ure, Heckman:2020otd,Baume:2021qho,
Distler:2022yse, Heckman:2022suy} as well as \cite{Heckman:2018jxk,
Argyres:2022mnu} for reviews of recent progress.}

In spite of their seemingly exotic higher-dimensional character, it is by now
well appreciated that the mere existence of these higher-dimensional field
theories provides important insights into a host of questions in quantum field
theory in diverse dimensions. One of the challenges in developing this
beautiful picture further is that in all known examples, there is no known
formulation in terms of a weakly coupled Lagrangian.  This in turn makes it
more difficult to extract ``microscopic details'' of the operator content and
correlation functions, which would no doubt provide a far better understanding
of many phenomena in the study of quantum fields at strong coupling, although
progress in that direction has been made in the context of the superconformal
bootstrap \cite{Beem:2015aoa, Chang:2017xmr, Baume:2021chx}.

Now, even though there is strictly speaking no weakly coupled limit, it is
still possible to sometimes perform a perturbative expansion in a small
parameter by restricting to suitable subsectors of the theory. This technique was used to
spectacular effect in the planar limit of $\mathcal{N} = 4$ super Yang--Mills at large 't Hooft coupling, where it was found that certain large
R-charge operators can be understood in terms of an integrable Heisenberg spin
chain \cite{Berenstein:2002jq}, which actually extends to the full
superconformal algebra \cite{Minahan:2002ve, Beisert:2003jj, Beisert:2003yb,
Beisert:2003xu, Beisert:2003jb, Beisert:2003ys}. More
broadly, the use of large R-charge expansions has been used to gain remarkable
insights into many strongly coupled systems \cite{Hellerman:2015nra,
Hellerman:2017sur, Alvarez-Gaume:2019biu, Hellerman:2020sqj, Hellerman:2021yqz,
Hellerman:2021duh}.

Our aim in this paper is to carry out a large R-charge expansion for a
specific class of quiver-like 6D SCFTs. Following the analysis presented in
\cite{Baume:2020ure, Heckman:2020otd}, we consider 6D SCFTs given by $L$
M5-branes probing an ADE singularity. Many other 6D SCFTs can be reached from
this starting point by Higgs branch or tensor branch renormalization group flows
\cite{Heckman:2018pqx}, so in this sense they provide us with access to a
ubiquitous class of operators. As found in \cite{DelZotto:2014hpa}, these
theories have a partial tensor branch given by a generalized quiver gauge
theory:
\begin{equation}\label{partialtensor}
		[G_{0}] - G_{1} - \cdots - G_{L-1} - [G_{L}] \,.
\end{equation}
Each $G_{i} = G$ is of the same ADE type, with $G_{0}$ and $G_{L}$
corresponding to flavor symmetries, and the rest corresponding to gauge
symmetries. Each vector multiplet is accompanied by a tensor multiplet, as
required to satisfy 6D anomaly cancellation constraints. The links
between $(G_{i-1},G_{i})$ for $i=1,\cdots,L$ are specified by ``conformal
matter'' \cite{DelZotto:2014hpa,Heckman:2014qba} (see also
\cite{Bergman:2020bvi}), a generalization of a weakly coupled hypermultiplet as associated with the edge modes localized on a single M5-brane probing an ADE
singularity.

There is by now substantial evidence that in the $[G_{i-1}]-[G_{i}]$
conformal matter theory, there are scalar half-BPS operators $X_{i}$ in the bifundamental
representation with scaling dimension $2,4,6,8,12$ in the case of
$G=A_{k},D_{k},E_{6},E_{7},E_{8}$ conformal matter, respectively \cite{Heckman:2014qba}. Our
operating assumption throughout this work will be that such conformal matter
operators do indeed exist, and that they constitute a well-defined
generalization of the standard bifundamental hypermultiplet. The case of the
partial tensor branch theory in equation \eqref{partialtensor} can then be obtained by
weakly gauging the associated flavor symmetries of conformal matter. As found
in \cite{Heckman:2014qba}, there is then a corresponding rank $L$
generalization of conformal matter with operators of respective dimensions $2L, 4L, 6L, 8L,
12L$ for rank $L$ conformal matter of type $G=A_{k},D_{k},E_{6},E_{7},E_{8}$. Such operators are composed out of the conformal matter operators of each conformal matter link as
\begin{equation}\label{def-X^L}
	X^{L} \equiv X_{1} X_{2} \cdots  X_{L} \,.
\end{equation}
The operator $X_{i}$ at each link transforms in a representation of the $\mathfrak{su}(2)_R$
R-symmetry of the 6D SCFT, and as such, we can consider other gauge invariant operators as
associated with the other conformal matter operators in the same R-symmetry
representation. In \cite{Baume:2020ure, Heckman:2020otd}, this was used to
construct a class of large R-charge operators of the general form:
\begin{equation}
	\mathcal{O}^{(m_{1},\cdots ,m_{L})} = X^{(m_{1})}_{1} \cdots  X^{(m_{L})}_{L} \,,
\end{equation}
where $m_{i} = -s,\cdots ,s$ denotes the Cartan charge associated with a spin $s$
representation of the $\mathfrak{su}(2)_{R}$ R-symmetry. When nearly all the
$m_{i}$ are highest weight states, references \cite{Baume:2020ure,
Heckman:2020otd} found that the operators $\mathcal{O}^{(m_{1},\cdots ,m_{L})}$ have
one-loop operator mixing governed by an \emph{integrable} XXX$_{s = \frac{j}{2}}$
Heisenberg spin chain with nearest neighbor interaction terms, where $j=2s$ is
the largest Dynkin label of the corresponding $\mathfrak{su}(2)_R$
representation. In this correspondence, the Hamiltonian $\mathcal{H}_{1D}$ of
the associated one-dimensional spin chain directly corresponds to the
dilatation operator $\mathcal{D}_{6D}$ acting on this operator subsector.

In the case of $j = 1$, this can be verified by a direct computation which
uses the holographic dual as a natural regulator to specify a fixed (albeit
large) value of the gauge coupling. Importantly, even though this operator
mixing is performed at strong coupling in the associated SCFT with the gauge
coupling scaling as $g_{YM}^{2} \sim L^{2/3}$, operator mixing is also
controlled by an inverse R-charge expansion going as $1/L^{2}$, so the
resulting operator scaling dimensions end up organizing according to a large
R-charge expansion:
\begin{equation}
	\Delta= \Delta_{0} + \frac{O(L^{2/3})}{L^{2}} + \cdots \,,
\end{equation}
where the ``order one coefficient'' can be computed using the 5D Kaluza--Klein-regulated theory obtained from placing the 6D theory on a large $S^1$.

Of course, once we assume the existence of conformal matter operators, a number
of additional consequences follow, just from superconformal symmetry.  For
example, since rank one $(G,G)$ conformal matter defines a generalization of a
half-BPS hypermultiplet, we also get an entire tower of (super)conformal
descendants. Denoting one such descendant as $\Phi^{(\ell_{i})}_{i}$ for
$(G_{i-1}, G_{i})$ conformal matter with $\ell_{i}$ an $\mathfrak{osp}(6,2 |
1)$ quantum number, we can then construct a wide variety of additional gauge
invariant operators of the full quiver:
\begin{equation}\label{confmatt}
	\Phi^{(\ell_{1})}_{1} \Phi^{(\ell_{2})}_{2} \cdots  \Phi^{(\ell_{L})}_{L}\,.
\end{equation}
Said differently, the $\mathfrak{su}(2)_{R}$ R-symmetry is simply a subalgebra
of the full superconformal algebra $\mathfrak{osp}(6,2| 1)$ of a $\mathcal{N} =
(1,0)$ SCFT. From this perspective, it is natural to study the operator content
and mixing associated with this larger super-spin chain.

In this work, we show that to leading order in a large R-charge expansion,
operator mixing for the operators in equation \eqref{confmatt} organizes
according to an integrable $\mathfrak{osp}(6,2|1)$ super-spin chain. To
establish this, we show that operator mixing of the full super-spin chain is
essentially controlled by the interactions governing the $\mathfrak{su}(2)_R$
R-symmetry spin chain through superconformal symmetry. Indeed, it turns out
that these diagrams are essentially the same as the ones associated with the
closely related 4D $\mathcal{N} = 2$ SCFT obtained from compactification on a
$T^{2}$, where we can directly verify both explicitly in the field theory, and
via a chain of string dualities (as associated with the string construction of
the SCFT). For examples of other spin chains in the context of 4D $\mathcal{N} = 2$ SCFTs, see e.g.,
\cite{Baume:2020ure, Heckman:2020otd,Beisert:2005he, Gadde:2009dj, Gadde:2010zi, Rey:2010ry, Liendo:2011xb,
Pomoni:2011jj, Gadde:2012rv, Pomoni:2013poa, Pomoni:2019oib, Pomoni:2021pbj, Gaberdiel:2022iot}.

With this in place, our task reduces to explicitly determining the operator
content associated with this super-spin chain. This provides us with direct
access to the microscopic content of the spin chain, as obtained from the
individual (super)conformal descendants of a hypermultiplet (in the case of
$(A,A)$ conformal matter) and its conformal matter generalizations (in the
case of $(D,D)$ and $(E,E)$ conformal matter). We emphasize that with
relatively mild assumptions, this is far more detailed information than what
would be obtained by just labeling states according to a representation of
the overall $\mathfrak{osp}(6,2|1)$ superconformal algebra.

The Bethe ansatz provides a general method for extracting the energy spectrum
for the corresponding integrable super-spin chain. The essential point is that once
we specify a ``ground state'' operator, namely the operator $X^{L}$ defined in
equation \eqref{def-X^L}, excitations in which we swap out some of the $X_{i}$
for superconformal descendants $\Phi_{i}^{(\ell_{i})}$ are governed by a
general Bethe ansatz for the spin chain with open boundary conditions of the
schematic form:
\begin{equation}\label{eqn:bethe2}
	\left(  \frac{u_{j,J}+\frac{i}{2}w_{J}}{u_{j,J}-\frac{i}{2}w_{J}}\right)^{2L}
	=
	\prod_{K=1}^{r}\prod_{\substack{k=1\\(k,K)\neq(j,J)}}^{n^{K}}
	\frac{u_{j,J}-u_{k,K}+\frac{i}{2}\mathcal{A}_{JK}}{u_{j,J}-u_{k,K}-\frac{i}{2}\mathcal{A}_{JK}}\,
	\frac{u_{j,J}+u_{k,K}+\frac{i}{2}\mathcal{A}_{JK}}{u_{j,J}+u_{k,K}-\frac{i}{2}\mathcal{A}_{JK}}\,,
\end{equation}
where, $n^K$ is the number of each type of excitation, the $u_{j,J}$
denote rapidities for each type of excitation, the $w$ indicate the weights of
the highest weight state for a representation of $\mathfrak{osp}(6,2|1)$, and
$\mathcal{A}$ is the symmetrized Cartan matrix defining a bilinear pairing on
the root system associated with the Dynkin diagram of the superconformal
algebra. The scaling dimensions/eigenenergies are then given in terms of the
rapidities by:
\begin{equation}
		(\Delta - \Delta_{0}) = E = \lambda_{G} \left\vert \sum_{J=1}^r\sum_{j=1}^{n^J} \frac{i}{u_{j,J} + \frac{i}{2}w_{J}} - \frac{i}{u_{j,J} - \frac{i}{2}w_{J}} \right\vert\,,
\end{equation}
where $\Delta_{0}$ denotes the scaling dimension in the ``free field limit'',
i.e., the scaling dimension of the tensor branch theory in which we neglect all
interaction terms. The constant $\lambda_{G}$ is a calculable coefficient
which depends on $G$, the ADE gauge group, as well as $L$, the length of the quiver.

As noted in reference \cite{Beisert:2003yb}, for Lie superalgebras and their
associated Dynkin diagrams, there are actually several distinct Dynkin diagrams
associated with the same Lie superalgebra. In the context of the Bethe ansatz,
each of these diagrams corresponds to a specific choice of ground state and
lowest energy excitations, i.e., each one gives us access to different
realizations of the same set of operators. Another aim of our analysis will be
to collect the explicit super-Dynkin diagrams as well as their corresponding
Bethe ansatz equations.

With the explicit Bethe ansatz in hand, we can also extract specific closed
operator subsectors which do not mix with other operators of the super-spin
chain. Examples include the $\mathfrak{su}(2)_{R}$ excitations as obtained by
just taking the original highest weight states $X_{i}$ and rotating to more
general $\mathfrak{su}(2)_{R}$ quantum numbers. In the holographic dual with
geometry $\text{AdS}_{7} \times S^{4} / \Gamma_{ADE}$, these correspond to small
excitations in the $S^{4} / \Gamma_{ADE}$ directions. Closely related to this
is the set of operators obtained by light-cone covariant derivatives with
respect to $G_{i-1}\times G_i$ acting on $X_{i}$, namely $D X_{i}$. In this
case, there is a corresponding ``$\mathfrak{sl}(2,\mathbb{R})$'' spin chain
subsector as associated with excitations in the $\text{AdS}_{7}$ direction. We also
find some additional closed subsectors, such as those associated with fermionic excitations.

We primarily focus on 6D SCFTs with $(A,A)$ conformal matter, but the analysis
we present readily generalizes to closely related systems.  For example, we can
obtain a little string theory by a diagonal gauging of the $G_0 \times G_L$
flavor symmetry of the 6D SCFT whose tensor branch is described in equation \eqref{partialtensor} \cite{Bhardwaj:2015oru}, and the only impact on
the spin chain is a change to periodic boundary conditions. Similarly, for 6D
and 4D theories with conformal matter, the only change is in the specification
of the highest weights $w_{I}$ of the spin chain excitations. An additional
comment here is that our considerations provide additional details on how
$(G,G)$ conformal matter, including its (super)conformal descendants generalize
the standard notion of a weakly-coupled hypermultiplet.

The rest of this paper is organized as follows. We begin in section
\ref{sec:REVIEW} by reviewing operator mixing in the $\mathfrak{su}(2)_{R}$
subsector of rank $L$ conformal matter, including its geometric interpretation
in a string compactification. Section \ref{sec:DYNKIN} determines a convenient
Dynkin diagram for $\mathfrak{osp}(6,2|1)$, as well as the preferred choice
where excitations of the spin chain coincide with superconformal descendants of
half-BPS conformal matter. Section \ref{sec:INTEGRO} establishes integrability
for operator mixing in this large R-charge sector for A-type conformal matter.
We turn to generalizations in Section \ref{sec:GEN}. Section \ref{sec:BETHE}
explores the application of the Bethe ansatz for these super-spin chains. We
present conclusions and future directions in Section \ref{sec:CONC}. In
Appendix \ref{app:oscillator-construction}, we present an oscillator
construction for 6D and 4D superconformal algebras. This appendix also collects
the Dynkin diagrams for the various superconformal algebras.

\section{The \texorpdfstring{\boldmath{$\mathfrak{su}(2)_{R}$}}{su(2)R} Spin Chains of Conformal Matter}\label{sec:REVIEW}

In this section, we present a brief review of the one-dimensional spin
chain describing an $\mathfrak{su}(2)_{R}$-sector of certain 6D SCFTs found in reference \cite{Baume:2020ure}. The 6D SCFTs we
consider are realized via a stack of M5-branes probing an ADE singularity. The geometry transverse to the M5-branes is given by $\mathbb{R}_{\bot}\times\mathbb{C}^{2}/\Gamma_{ADE}$, where the group
$\Gamma_{ADE}$ acts \textquotedblleft holomorphically\textquotedblright on the
coordinates $(u,v)$ of the $\mathbb{C}^{2}$ geometry. This leaves unbroken an
$\mathfrak{su}(2)$ R-symmetry, which acts upon the combination $(u,\overline{v})$.

Much of our understanding of these 6D SCFTs comes from the study of their partial tensor branch.
The partial tensor branch is obtained by keeping the M5-branes at the singularity,
but separated in the $ \mathbb{R}_{\bot}$ direction. Doing so, we arrive at a
6D generalized quiver gauge theory of the form:
\begin{equation}
	[G_{0}]-G_{1}-\cdots -G_{L-1}-[G_{L}] \,.
\end{equation}
Each of the links in this generalized quiver corresponds to the degrees of
freedom localized on a single M5-brane. In particular, references
\cite{DelZotto:2014hpa, Heckman:2014qba} demonstrated that there is a class of
scalar operators $X_{i}$ transforming in the bifundamental representation of
$(G_{i-1},G_{i})$ known as ``conformal matter'' operators, a generalization of the
$\mathfrak{su}(2)_{R}$ highest weight state scalars of the standard hypermultiplet.\footnote{By abuse of conventions, we often refer to ``conformal matter'' as also specifying a 6D SCFT \cite{DelZotto:2014hpa}. For example, in the special case where $G_{i}=SU(K)$ for each $i$, conformal matter just corresponds to a 6D hypermultiplet in the bifundamental representation of $(G_{i-1},G_{i})$.}
A natural class of half-BPS operators of this generalized quiver are obtained by
taking the products of these $X_{i}$, which we refer to as $X^{L}$:
\begin{equation}\label{defX^L}
		X^{L}=X_{1}X_{2}\cdots X_{L}\,,
\end{equation}
where we have implicitly contracted all neighboring gauge indices, and
suppressed the flavor indices.  These are half-BPS operators and as such are
expected to specify operators of the 6D SCFT in the limit where we return to
the origin of the tensor branch. Indeed, this type of operator can be
identified in terms of branes stretched from the north to the south pole of the
orbifold fixed points in the $\text{AdS}_7 \times S^4 / \Gamma_{ADE}$ holographic dual
\cite{Bergman:2020bvi, Baume:2020ure, Heckman:2020otd}. The exact scaling
dimension has been determined using methods from the associated holomorphic
F-theory geometry, as well as the holographic dual \cite{Heckman:2014qba,
Bergman:2020bvi, Baume:2020ure}.

However, this is only one possible gauge invariant operator which we can form on the
partial tensor branch. For example, in the limit where we ``ungauge'' the
$(G_{i-1},G_{i})$ symmetry, we observe that conformal matter operators specify
a representation of the superconformal algebra in their own right. As such, we
can consider other related operators obtained from $\mathfrak{su}(2)_{R}$
R-symmetry transformations, or more general superconformal transformations.
Phrased in this way, we can speak of a ``ground state'' for a spin chain, as
obtained by taking the tensor product of the vacuum of a single spin $L$
times:
\begin{equation}
	\left|\text{GND}\right> =\left|0\right>_{1}\otimes\cdots\otimes\left| 0\right>_{L}\,.
\end{equation}
Viewing each $\left\vert 0\right\rangle _{i}$ as the highest weight state of an
$\mathfrak{su}(2)_{R}$ R-symmetry representation, we can build up more general
states by acting with the $\mathfrak{su}(2)$ lowering operator.

Specializing to the case of the hypermultiplet, the associated
$\mathfrak{su}(2)_R$ states fill out a collection of doublets (one for each
link), which we can write as $(X_{i},Y_{i}^{\dag})$, related to the vacuum by
\begin{equation}
		X_{i}\longleftrightarrow \left|0\right>_i\,,\qquad Y^\dagger_i\longleftrightarrow J^-\left|0\right>_i\,,
\end{equation}
Now, as opposed to the case of $X^{L}$, an operator such as:
\begin{equation}
	X^{M}Y^{\dag}X^{N}Y^{\dag}X^{L-M-N} =
	X_{1}X_{2}\cdots X_{M}Y_{M+1}^{\dag}X_{M+2}X_{M+3}\cdots X_{M+N+1}Y_{M+N+2}^{\dag}X_{M+N+3}\cdots X_{L}\,,
\end{equation}
will not generically be half-BPS. These operators are in one-to-one correspondence with the states of an
$\mathfrak{su}(2)_R$ spin chain. Indeed, each $X_{i}$ and $Y_{i}^{\dag}$
respectively specify spin up and spin down excitations.

Motivated by similar structures appearing in \cite{Berenstein:2002zw}, it is
natural to expect that operator mixing for this special class of operators is
controlled by a spin chain Hamiltonian with nearest neighbor interactions. This
can be confirmed by a direct calculation of operator mixing in the regime where
the tensor multiplet vevs $\left\langle t_{i}\right\rangle =1/g_{i}^{2}$
specify energy scales high above that of the effective field theory, and
similar considerations hold for the 4D $\mathcal{N}=2$ SCFTs obtained by
compactification of the partial tensor branch theory \cite{Baume:2020ure}.
Based on this, it is tempting to extrapolate this behavior to the conformal
fixed point, but we face the technical challenge that precisely at this point,
the gauge couplings are formally infinite. The situation is not entirely
hopeless, however, because this class of 6D SCFTs have holographic duals
given by the geometry $\text{AdS}_{7}\times S^{4}/\Gamma$, with orbifold fixed points
at the north and south poles. Using the
holographic dual description, we get a large value of each gauge coupling in
the quiver which scales as $L^{2/3}$, with $L$ the number of M5-branes. For
operators with large R-charge, however, the strength of operator mixing for
spin chain operators is naturally suppressed by an overall $1/L^{2}$ factor, so, at least for operators with a low number of impurities, there is a systematic
inverse R-charge expansion available, and perturbation theory remains valid.
Upon diagonalization of the dilatation operator, this can all be summarized in
terms of eigenoperators with scaling dimension:
\begin{equation}
	\Delta=\Delta_{0}+\frac{\alpha}{L^{2}}+\cdots.
\end{equation}
Here, $\Delta_{0}$ denotes the scaling dimension of the half-BPS conformal matter operator of the spin chain:
\begin{equation}\label{delta0}
\Delta_{0} = L \Delta_{X} \,,
\end{equation}
where $\Delta_{X}$ denotes the scaling dimension of the bifundamental minimal conformal
matter operator. The different values of $\Delta_{X}$ for 6D and 4D SCFTs known as conformal matter of type $(G,G)$ are as follows:
\begin{equation}\label{taboo}
	\begin{tabular}
	[c]{|c|c|c|c|c|c|}\hline
	& $SU(K)$ & $SO(2K)$ & $E_{6}$ & $E_{7}$ & $E_{8}$\\\hline
	$w$ & $1$ & $2$ & $3$ & $4$ & $6$\\\hline
	$\Delta_{X}^{6D}$ & $2$ & $4$ & $6$ & $8$ & $12$\\\hline
	$\Delta_{X}^{4D}$ & $1$ & $2$ & $3$ & $4$ & $6$\\\hline
	\end{tabular}
\end{equation}
and in the first line we have also indicated the integer-normalized highest
weight $w$ of the relevant $\mathfrak{su}(2)_{R}$ representation. Note that the conformal
dimension of these half-BPS states is set by their R-charge,
$\Delta_X=\frac{D-2}{2}w$ as expected from shortening conditions in
superconformal representation theory \cite{Dolan:2002zh, Buican:2016hpb,
Cordova:2016emh}.

As found in \cite{Baume:2020ure}, in the limit where the number of impurity
insertions $I\ll L$, all of the operator mixing for $(G,G)$-type conformal
matter is captured by a Hamiltonian with nearest neighbor
interactions:\footnote{Compared with \cite{Baume:2020ure, Heckman:2020otd} and
\cite{Faddeev:1996iy}, we use the mathematics convention and have
integer R-charges, $ j_\text{here} = 2s_\text{there}$, for all
$\mathfrak{su}(2)$ representations so that a ``qubit'' with a spin up
and spin down state will be labeled by $j = 1$. This will be more
convenient when discussing the full superconformal algebra. Moreover,
the length of the spin chain will be denoted by the number of sites,
rather than the number of gauge/flavor groups: $L_\text{here} =
(N+1)_\text{there}$.}
\begin{equation}\label{hamiltonian-su2}
	H=-\lambda_{G}\underset{i}{\sum}Q_{w}(\overrightarrow{S}_{i}\cdot\overrightarrow{S}_{i+1})\,,
\end{equation}
where the value of $w$ and $G$ are determined by the type of conformal matter
we are considering. In the above, $Q_{w}(x)$ is a specific degree $w$
polynomial (see e.g., \cite{Faddeev:1996iy}), and the value of the coupling
$\lambda_{G}$ is:
\begin{equation}
	\lambda_{G}=\frac{(\pi L)^{2/3}}{16\pi^{3}}\times\widetilde{C}_{G}\,,
\end{equation}
with $\widetilde{C}_{G}$ specified by purely group theoretic data:
\begin{equation}\label{lambda_G}
	\widetilde{C}_{SU(K)} = \frac{K^{2}-1}{2K^{2}}\,,\quad
	\widetilde{C}_{SO(2K)} = \frac{2K-1}{2K-2}\,,\quad
	\widetilde{C}_{E_{6}}=\frac{13}{18}\,,\quad
	\widetilde{C}_{E_{7}}=\frac{19}{24}\,,\quad
	\widetilde{C}_{E_{8}}=1 \,.
\end{equation}

At a qualitative level, the appearance of nearest neighbor interactions follows
directly from locality in the holographic dual $\text{AdS}_7 \times S^4 /
\Gamma_{ADE}$, where excitations amount to local fluctuations along the great
arc connecting the orbifold fixed points at the north and south poles. An
important feature of the quiver description is that the spine of the quiver /
spin-chain is directly specified by this great arc. As such, the holographic
dual determines a regulated value for the gauge coupling. A direct calculation
of hopping can also be carried out in the case of $(A,A)$ conformal matter. An
important technical element of this calculation is to work with respect to the
5D Kaluza--Klein regulated theory \cite{Baume:2020ure}. In this procedure, one treats all
fields of the 6D theory as 5D fields, but in which one integrates over all six
spacetime dimensions. An a posteriori justification for this procedure is that
in calculating the corresponding loop diagrams in the KK regulated theory, one
finds logarithmic corrections to operator mixing which precisely match the
expected hopping structure. We refer the reader to \cite{Baume:2020ure} for additional details.

Now, although the individual Hamiltonians are quite involved, integrability
essentially fixes a unique choice, and this can be solved via the algebraic
Bethe ansatz method (see e.g., \cite{Bethe:1931} and \cite{Faddeev:1996iy} for
a review). The end result for the open spin chain is that excitations are
controlled by complex rapidities $u_{i}$ which are related to the momenta of
the excitations via:
\begin{equation}
	\exp(ip_{j})=\frac{u_{j}+i\frac{w}{2}}{u_{j}-i\frac{w}{2}} \,,
\end{equation}
and these satisfy the Bethe ansatz equations for a spin chain with $L$ sites, and open
boundary conditions (see, e.g., \cite{Cao:2013qxa, WANG}):
\begin{equation}
	\left(  \frac{u_{j}+i\frac{w}{2}}{u_{j}-i\frac{w}{2}}\right)  ^{2L}= \underset{l\neq j}{{\displaystyle\prod}}
	\frac{\left(  u_{j}-u_{l}+i\right)  }{\left(  u_{j}-u_{l}-i\right)  }
	\frac{\left(  u_{j}+u_{l}+i\right)  }{\left(  u_{j}+u_{l}-i\right)  }\,.
\end{equation}

To apply this in the study of operators in 6D SCFTs, we must also require that
the net momentum along the spin chain vanishes, so that it does not leak off
the ends of the M5-brane probe theory \cite{Baume:2020ure}. This amounts to the
following condition on the rapidities:
\begin{equation}\label{eqn:conserve}
	\underset{j}{{\displaystyle\prod}}
	\frac{u_{j}+i \frac{w}{2}}{u_{j}-i \frac{w}{2}}=1\,.
\end{equation}
In the case of 4D SCFTs obtained from compactification of the partial tensor branch theory on a $T^2$, this constraint is absent since momentum can instead dissipate along the extra-dimensional $T^2$ direction.

With the rapidities in hand, we can then proceed to the determination of the energy of a given eigenoperator.
For rank $L$ conformal matter of type $(G,G)$, this is given by:
\begin{equation}\label{eqn:absofacto}
		(\Delta - \Delta_{0}) = E_{G}= \lambda_{G} \left\vert \sum_{j}\epsilon(u_j) \right\vert\,,
\end{equation}
where $\Delta_{0}$ specified as in equations \eqref{delta0} and \eqref{taboo}, and the energy of a given quasi-particle excitation is:
\begin{equation}
\epsilon(u) = \frac{i}{u + i\frac{w}{2}} - \frac{i}{u - i\frac{w}{2}} \,.
\end{equation}
The presence of an absolute value in equation (\ref{eqn:absofacto}) is the physical requirement that we select positive energy excitations above the ground state.

Similar considerations hold for little string theories (LSTs). Starting from rank $L$ 6D conformal matter of type $(G,G)$, we can gauge a common diagonal flavor symmetry and add an additional tensor multiplet with charge $-2$ to cancel the corresponding gauge anomalies. This results in a circular quiver, and in the spin chain it just means we now have periodic boundary conditions. The Bethe ansatz equations in this case are:
\begin{equation}
	\left(  \frac{u_{j}+i\frac{w}{2}}{u_{j}-i\frac{w}{2}}\right)  ^{L}= \underset{l\neq j}{{\displaystyle\prod}}
	\frac{\left(  u_{j}-u_{l}+i\right)  }{\left(  u_{j}-u_{l}-i\right)  }
	\,,
\end{equation}
and we again must impose a momentum conservation constraint as in equation (\ref{eqn:conserve}).
Let us note that 4D $\mathcal{N} = 2$ SCFTs obtained from circular
quiver gauge theories with A-type gauge groups were studied in \cite{Beisert:2005he},
though the analysis of \cite{Baume:2020ure} generalizes this to the case
of quivers with 4D conformal matter.

\subsection{Geometric Interpretation}

Though our primary focus will be on the representation theoretic structure
associated with conformal matter operators, it is natural to ask whether these
``nearly BPS'' structures can be identified in the corresponding string
compactification geometry. Following \cite{Heckman:2014qba}, we recall that the local
singularity associated with rank one ADE conformal matter takes the form:
\begin{align}
(A_{k}, A_{k}): y^{2}  &  = x^{2} + (uv)^{k+1}\\
(D_{p}, D_{p}): y^{2}  &  = (uv)x^{2} + (uv)^{p-1}\\
(E_{6}, E_{6}): y^{2}  &  = x^{3} + (uv)^{4}\\
(E_{7}, E_{7}): y^{2}  &  = x^{3} + (uv)^{3}x\\
(E_{8}, E_{8}): y^{2}  &  = x^{3} + (uv)^{5}.
\end{align}
To get rank $L$ conformal matter, we consider the orbifold group action $(u,v)
\mapsto(\zeta u, \zeta^{-1} v)$ with $\zeta$ a primitive $L$th root of
unity, namely $\zeta^{L} = 1$. In the corresponding F-theory geometry, this can
be visualized as a collision of $G$-type seven-branes intersecting at the
orbifold fixed point $\mathbb{C}^{2} / \mathbb{Z}_{L}$. As found in
\cite{Heckman:2014qba}, there is a special class of complex structure
deformations associated with 7-brane recombination, given by $uv \mapsto uv - r$.
This specifies a familiar correspondence in which vacuum expectation values of
holomorphic operators directly translate to complex structure deformations. For
example, the same phenomenon also applies in 4D $\mathcal{N} = 2$ theories to
the Coulomb branch of the associated Seiberg--Witten geometries.

There is a natural correspondence between non-harmonic representatives of the
deformation ring, and the operators of the $\mathfrak{su}(2)_{R}$ spin chain. To
see why, observe first that the polynomial $u^{L}$ is invariant under the
$\mathbb{Z}_{L}$ group action, and this directly corresponds to the $X^{L}$
spin chain operator. Similarly, the polynomial $v^{L}$ maps to the operator
$Y^{L}$. Next, note that the doublet $(u , \overline{v})$ transforms in the
same R-symmetry representation as $(X,Y^{\dag})$. Consequently, we can also
build polynomials $u^{M} \overline{v}^{L-M}$, which, loosely speaking, map to
spin chain states with $M$ insertions involving the $X$ conformal matter
operators, and $L-M$ insertions involving the $Y^{\dag}$ conformal operators.
It is at this point that the correspondence with geometry appears to break
down, because whereas there are many different words we can form in the spin
chain (depending on the precise ordering of the $X$ and $Y^{\dag}$ insertions),
in the commutative ring there is just a single (non-holomorphic) polynomial
$u^{M} \overline{v}^{L-M}$.

In fact, more deformations are available, and can be seen by considering the
resolution of the $\mathbb{C}^{2} / \mathbb{Z}_{L}$ singularity. Performing
the resolution, we get a collection of $L-1$ compact $\mathbb{P}^{1}$s,
each with homogeneous coordinates $[u_i, v_i]$ for $i = 0,\cdots,L$, where the case
$i = 0$ and $i = L$ correspond to flavor brane curves which are actually
non-compact. In the vicinity of each collision, we have the rank one conformal
matter:
\begin{align}
	(A_{k}, A_{k}): y^{2}  &  = x^{2} + (u_{i}v_{i+1})^{k+1}\\
	(D_{p}, D_{p}): y^{2}  &  = (u_{i}v_{i+1})x^{2} + (u_{i}v_{i+1})^{p-1}\\
	(E_{6}, E_{6}): y^{2}  &  = x^{3} + (u_{i}v_{i+1})^{4}\\
	(E_{7}, E_{7}): y^{2}  &  = x^{3} + (u_{i}v_{i+1})^{3}x\\
	(E_{8}, E_{8}): y^{2}  &  = x^{3} + (u_{i}v_{i+1})^{5} \,.
\end{align}
The point is that we can now make the identification between $u^{L}$ and the
weight $L$ polynomial $u_{1}\cdots u_{L}$, and $v^{L}$ and $v_{0} \cdots
v_{L-1}$. In this case, starting from the single $\mathbb{Z}_{L}$ monomial
$u^{M} \overline{v}^{L-M}$, we again see that there are just as many ways to
build a polynomial out of the $u_{i}$ and $\overline{v}_{j}$ as there are spin
chain states.  There is a natural sense in which the geometric R-symmetry which
acts on the doublet $(u,\overline{v})$ extends to an $\mathfrak{su}(2)$ action
on each set of local coordinates $(u_{i}, \overline{v}_{i+1})$.

The vacuum expectation value of conformal matter corresponds to a complex
structure deformation of the corresponding Calabi--Yau.  These operators also
come with a spacetime position dependence, e.g., $u \mapsto u(x)$. Doing so, we
see that just as the R-symmetry implements a rotation in the internal
dimensions, there is a corresponding set of rotations and translations
available in the 6D spacetime. In the holographic dual, these are associated
with fluctuations in the $\text{AdS}_7$ direction.  More generally, viewing the
gravity dual as a super-coset, motion in the fermionic directions corresponds to
fermionic excitations of the spin chain. Part of our task will be to determine
the precise structure of the spin chain in the presence of these additional
types of impurities.

Finally, the natural expectation is that operator mixing of the spin chain
thus defines a quantum deformation of the original classical geometry. It
would be interesting to carry out such an analysis in detail, but it lies
outside the scope of the present work.

\subsection{Higher Symmetries Interpretation}

\begin{figure}[t!]
\begin{center}
\includegraphics[scale = 0.5, trim = {0cm 4.0cm 0cm 5.0cm}]{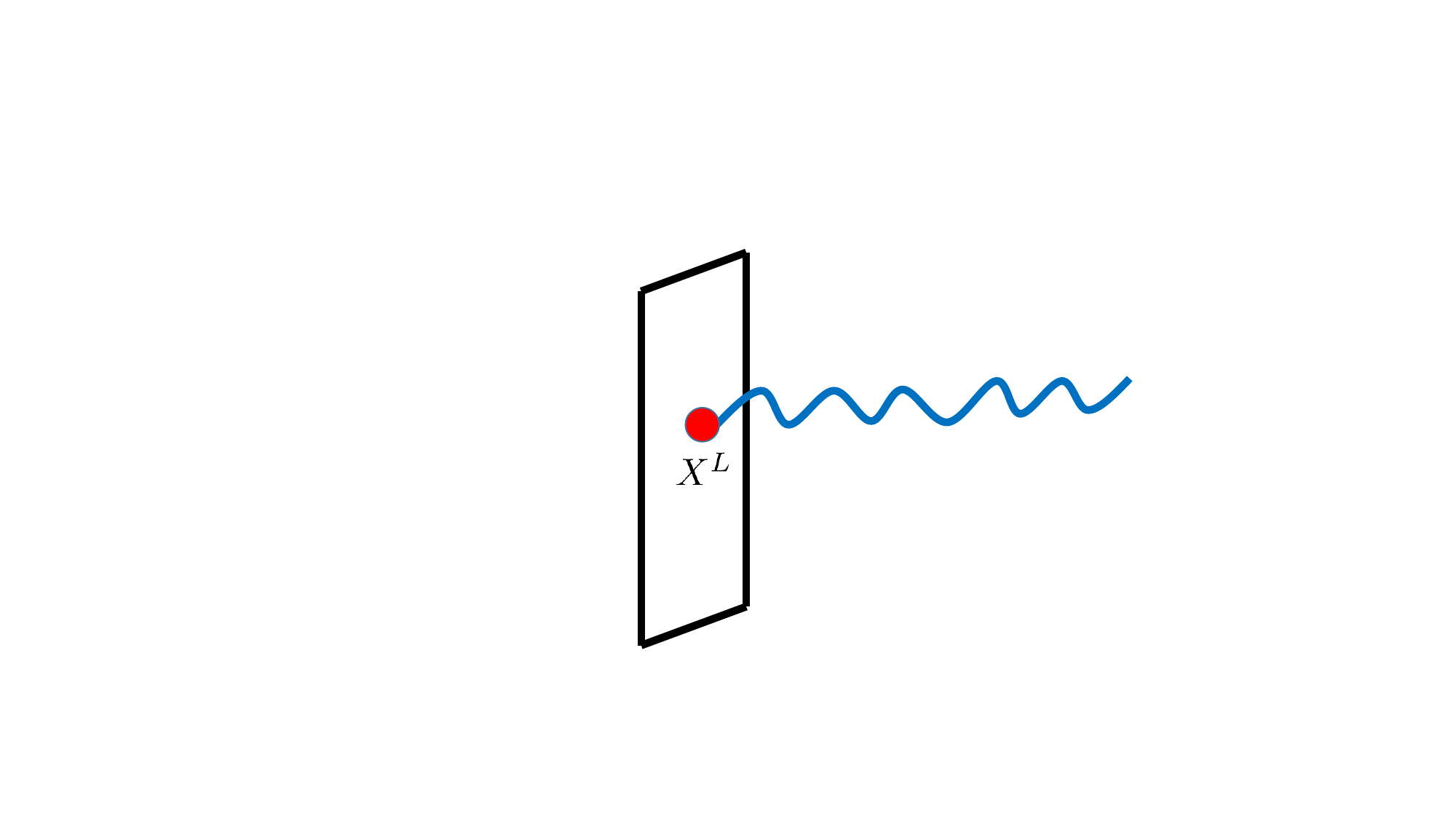}
\caption{Depiction of a Wilson line of 7D super Yang--Mills theory ending on a stack of M5-branes. The endpoint of the
Wilson line is given by a conformal matter operator $X^L$ of the corresponding 6D SCFT. Other choices of $\mathfrak{osp}(6,2|1)$
spin chain excitations specify non-supersymmetric boundary conditions for 7D Wilson lines ending on the defect.}
\label{fig:WilsonLine}
\end{center}
\end{figure}

Recently, there has been much progress in understanding various aspects of
higher symmetries in higher-dimensional QFTs (see e.g., \cite{DelZotto:2015isa,Heckman:2017uxe,Eckhard:2019jgg,GarciaEtxebarria:2019caf,Albertini:2020mdx,Morrison:2020ool,Dierigl:2020myk,Closset:2020scj,DelZotto:2020esg,Apruzzi:2020zot,Bhardwaj:2020phs,Closset:2020afy,Heidenreich:2020pkc,DelZotto:2020sop,Gukov:2020btk,Bah:2020uev,Bhardwaj:2021pfz,Apruzzi:2021mlh,Apruzzi:2021phx,Hosseini:2021ged,Apruzzi:2021vcu,Bhardwaj:2021wif,Bhardwaj:2021zrt,Closset:2021lwy,Heidenreich:2021xpr,Buican:2021xhs,Cvetic:2021maf,Apruzzi:2021nmk,Braun:2021sex,Bah:2021brs,Bhardwaj:2021mzl,Cvetic:2020kuw,Cvetic:2021sxm,Debray:2021vob, DelZotto:2022fnw,
Hubner:2022kxr,Cvetic:2022imb,DelZotto:2022joo,DelZotto:2022ras,Heckman:2022suy}
for recent work in this direction). In particular, in reference
\cite{Heckman:2022suy}, these structures were used to extract the global form
of the (continuous) zero-form symmetry for a broad class of 6D SCFTs, including
those obtained from M5-branes probing an ADE singularity. In the case of rank
$L$ conformal matter of type $(G,G)$, the global form is given by $G \times G /
\mathcal{Z}_{\mathrm{diag}}$.\footnote{The full quotient can also act on
various Abelian symmetry factors as well as the R-symmetry and spacetime
symmetries, see \cite{Heckman:2022suy} for details.} Now, one of the elements
in this construction is the construction of gauge invariant operators (on the
full tensor branch) which transform in the bifundamental representation of the
flavor group. On the partial tensor branch, we also have the closely related
half-BPS operators $X^L$. The requirement that large gauge transformations of
the center of the gauge group leave each conformal matter operator invariant
correlates the centers of these gauge groups, as well as the flavor symmetry
factors.

From the perspective of the 7D super Yang--Mills theory specified by M-theory
on $\mathbb{C}^2 / \Gamma_{ADE}$, the M5-brane probes define a codimension one
defect which couples to discrete bulk-three form potentials as well as the bulk gauge field.
In particular, we can consider a Wilson line which ends on $X^L$.
Other choices of spin-chain states correspond to deformations of this basic
configuration, and amount to other non-supersymmetric boundary conditions for
Wilson lines. For example, an R-symmetry excitation corresponds to also
switching on contributions from the adjoint valued R-symmetry triplet of
scalars in the 7D Wilson line. See Figure \ref{fig:WilsonLine} for a depiction
of one such Wilson line.

\section{Super-Spins and Dynkin Diagrams}\label{sec:DYNKIN}

In the previous section, we reviewed how long quiver-like 6D SCFTs admit a
large R-charge subsector in which operator mixing is governed by the
$\mathfrak{su}(2)$ Heisenberg spin chain. In the 6D theory, this
$\mathfrak{su}(2)$ is identified with the R-symmetry of the theory, and the
specification of a highest weight state for conformal matter dictates the
corresponding representations on each spin chain site. Continuing in this vein,
we now ask how the full superconformal algebra acts on a spin chain site.
Indeed, in the special case of rank one conformal matter, we can build an
infinite-dimensional representation in which the ``ground
state'' $X$ serves as the highest weight state of some
superconformal representation. From this perspective, we can now generate a far
broader class of possible spin-chain excitations simply by acting with the
generators of the superconformal algebra, $\mathfrak{osp}(6,2|1)$. Our eventual
aim will be to study operator mixing for all such spin chain excitations. To
set the stage for this analysis, in this section we study the content of a
single super-spin excitation.

At a certain level, this is straightforward: ``all'' that is required is for us
to specify a highest weight state, and to then generate the corresponding
descendants by acting with the generators of the superconformal algebra.
Indeed, we already implicitly performed this task in the case of the
$\mathfrak{su}(2)$ spin chain, where we began with the highest weight state $X$
and then acted by the lowering operator $J^{-}$ to build up the other scalar
components of the $\mathfrak{su}(2)_{R}$ multiplet.

Now, for a Lie (super)algebra $\mathfrak{g}$ we can always work with respect to
a basis of the generators in terms of triplets $(E_{I}^{+},E_{I}^{-},H_{I})$
with $I=1,\cdots,r=\text{rank}(\mathfrak{g})$ satisfying the following the
Serre--Chevalley relations:\footnote{We refer to, e.g., references
\cite{Kac:1977em, Frappat:1987ix, Frappat:1996pb} for reviews of Lie
superalgebras.}
\begin{equation}\label{Serre--Chevalley}
	[H_{I},H_{J}]=0\,,\qquad
	[E_{I}^{+},E_{J}^{-}\}=\delta_{IJ}H_{I}\,,\qquad
	[H_{I},E_{J}^{\pm}]=\pm\mathcal{M}_{JI}E_{J}^{+}\,,\qquad\text{(no sums)}\,,
\end{equation}
where $\mathcal{M}_{IJ}$ is the Cartan matrix of $\mathfrak{g}$, and
anticommutators are used for two fermionic generators. Here, the $E_{I}^{+}$
and $E_{I}^{-}$ respectively denote raising and lowering operators, and the
$H_{I}$ generate the Cartan subalgebra. Any other generators can then be found
by taking (anti)commutators of $E^\pm_I$.

These triplets determine a root system where we associate a root vector, $\alpha$, to
each generator, $\mathcal{J}$: $[H_I, \mathcal{J}] = \alpha_I \mathcal{J}$.
Those of the $r$ raising operators, $E^+_I$, satisfying the Serre--Chevalley
relations \eqref{Serre--Chevalley} are the simple roots. The root system comes
equipped with a pairing $\left<\alpha,\beta\right> =
\alpha_I(\mathcal{A}^{-1})^{IJ}\beta_J$, where $\mathcal{A}$ is referred to as
the symmetrized Cartan matrix. If $\mathfrak{g}$ is compact and simply laced, both
matrices, $\mathcal{M}$ and $\mathcal{A}$, agree, while in other cases, there always exists a matrix, $\Lambda$, such that
$\mathcal{A}= \Lambda\mathcal{M}$. This corresponds to a rescaling of the
Cartan generators, $H_{I}\rightarrow(\Lambda^{-1})_{IJ}H_{J}$ or, equivalently, of
the ladder operators. The matrix $\mathcal{A}$ is precisely the one appearing
in the Bethe ansatz in equation \eqref{eqn:bethe2}.

Any state can then be written as a linear combination of eigenstates labeled
by their weights $\left| h_{1},\cdots,h_{r}\right>$ with:
\begin{equation}\label{cartan-labels}
	H_{I}\left|h_{1},\cdots,h_{r}\right> = h_{I}\left| h_{1},\cdots,h_{r}\right>\,.
\end{equation}
A highest weight state $\left\vert \Omega\right\rangle =\left\vert
w_{1},\cdots,w_{r}\right\rangle$ is defined as one which is annihilated by all
the raising operators, namely:
\begin{equation}
E_{I}^{+}\left\vert \Omega\right\rangle =0\,,\qquad H_{I}\left\vert
\Omega\right\rangle =w_{I}\left\vert \Omega\right\rangle \,,
\end{equation}
where the $w_{I}$ are the weights of this state. Starting from such a highest
weight state, we can of course start producing various descendants of the
superconformal algebra:
\begin{equation}\label{descendents}
		\prod_{I=1}^{\text{rank}(\mathfrak{g})}\left(E_{I}^{-}\right)^{n^{I}}\left\vert \Omega\right\rangle \,.
\end{equation}
The basis defined by the Serre--Chevalley relations in equation \eqref{Serre--Chevalley}
ensures that the changes in the weight vector after successive applications of
the lowering operators, $E_{I}^{-}$, are encoded in the Cartan matrix, e.g.,
$H_{I}E_{J}^{-}\left\vert \Omega\right\rangle
=(w_{I}-\mathcal{M}_{JI})E_{J}^{-}\left\vert \Omega\right\rangle$.

Let us now specialize to the Lie superalgebra associated with the 6D
superconformal algebra, $\mathfrak{osp}(6,2|1)$.\footnote{The R-symmetry of
$\mathfrak{osp}(6,2|\mathcal{N})$ is $\mathfrak{sp}(\mathcal{N})$, so that in
our convention $\mathfrak{sp}(1)_R=\mathfrak{su}(2)_R$.} In the study of SCFTs,
it is often convenient to first consider the states as specified by the bosonic
subalgebra $\mathfrak{so}(6,2)\times\mathfrak{su}(2)_R$, and then we can specify
quantum numbers according to the subalgebra
$\mathfrak{so}(2)\times\mathfrak{so}(6)\times\mathfrak{su}(2)_R$, namely a
scaling dimension, three Lorentz quantum numbers and an R-charge:
\begin{equation}\label{quantum-numbers-6D}
	\mathfrak{so}(2)\times\mathfrak{so}(6)\times\mathfrak{su}(2)_R\text{ weights:}\quad[\Delta;\ell_{1};\ell_{2};\ell_{3};j] \,.
\end{equation}

For our present purposes, however, it will actually prove more convenient to
single out the maximal compact subalgebra $\mathfrak{u}(4|1)$ instead. The
Cartan subalgebra of $\mathfrak{u}(4|1)$ is of course identical to that of
$\mathfrak{so}(2)\times\mathfrak{so}(6)\times\mathfrak{su}(2)_R$, but it has
the advantage that unitary irreducible representations of this subalgebra are
finite-dimensional.  We can of course extend to the unitary
infinite-dimensional representations of the full $\mathfrak{osp}(6,2|1)$ by
acting with the remaining generators. In terms of the generators of the
superconformal algebra, this subalgebra is generated by:
\begin{equation}
	\mathfrak{u}(4|1)\text{ generators:}\quad
	\mathcal{L}_{j}^{i} \oplus J^{(3)} \oplus Q^{i+} \oplus S_{i-}\,,
\end{equation}
in the obvious notation, with $i,j$ indices in the fundamental (when raised)
and anti-fundamental (when lowered) of $\mathfrak{su}(4)=\mathfrak{so}(6)$. We
label the corresponding weights, consisting of the $\mathfrak{u}(4)$ generators
and the R-charge via:
\begin{equation}\label{u(4|1)-weights}
		\mathfrak{u}(4|1)\text{ weights:}\quad \vec{h} = (\ell_{1},\ell_{2},\ell_{3},h_4 ,j) \,.
\end{equation}
We stress that $h_4$ is related to $\Delta$ by a particular combination of the
other weights (see equation \eqref{eqn:DELTAFORCE} later on for the precise
relation).

Another complication we face is that in the case of Lie superalgebras, the
specification of a Dynkin diagram (and thus a collection of simple roots) is
not unique \cite{Kac:1977em, Frappat:1987ix}. Correspondingly, the
specification of highest weight states, as well as excitations above this
\textquotedblleft ground state\textquotedblright\ also depend on this
additional input. Indeed, for ordinary Lie algebras, each Weyl reflection
simply permutes the root system, and the Dynkin diagram remains invariant. For
Lie superalgebras, however, different super-Weyl reflections can result in
different Dynkin diagrams. To determine the full set of possible root systems,
we performed a brute force sweep over possible combinations of generators; The
results of this sweep are collected in Figure \ref{fig:dynkin-6d} of Appendix
\ref{app:oscillator-construction}.

\begin{figure}[t!]
		\centering
		{\resizebox{0.34\textwidth}{!}{\includegraphics{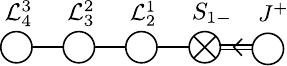}}}
		\caption{The ``Beauty'' Dynkin diagram of the 6D superconformal algebra
		$\mathfrak{osp}(6,2|1)$. Each node is labeled by the corresponding
		generator, $E^+_I$ satisfying the Serre--Chevalley relations
		of line \eqref{Serre--Chevalley}.}
		\label{fig:beauty-6d-main-text}
\end{figure}

There are two canonical choices of Dynkin diagram. One is the \textquotedblleft
distinguished Dynkin diagram\textquotedblright\ which extends the standard
bosonic subalgebra $\mathfrak{so}(6,2)\times \mathfrak{su}(2)$ by the minimum
number of additional fermionic roots. It is indicated in Figure
\ref{fig:distinguished_6D}. Additionally there is a Dynkin diagram in which a
given highest weight state also corresponds to a superconformal primary, namely
it is annihilated by the special conformal transformations $K$ as well as the
corresponding conformal supersymmetries $S$. The simple root system associated
with this choice is Figure \ref{fig:beauty-6d-main-text}, namely the only
fermionic nodes which appear involve $S$ (no $Q$ generators appear). Following
the nomenclature of reference \cite{Beisert:2003yb}, we refer to this choice
where highest weight states coincide with superconformal primaries as the
\textquotedblleft Beauty\textquotedblright\ and the distinguished Dynkin
diagram (where the full conformal symmetry is manifest)\ as the
\textquotedblleft Beast\textquotedblright.

As a last comment, in Appendix \ref{app:oscillator-construction} we provide
additional details on how we determined this choices by using an oscillator
construction of the corresponding superconformal algebra, as well as the
commutation relations between its generators. For completeness, we also give a
parallel treatment of the 4D\ $\mathcal{N}=2$ case.

To proceed further, we now discuss
in more detail representations of the superconformal algebra.

\subsection{The Beauty}\label{sec:beauty-6D}

As mentioned above, it will be most convenient to work in terms of a basis where
the highest weight states coincide with superconformal primaries. Since
$K_{ij}=\{S_{i+},S_{j-}\}$ and $S_{i+}\sim\lbrack J^{+},S_{i-}]$, states
annihilated by both $J^{+}$ and $S_{i-}$ are automatically superconformal
primaries. We are left with the Lorentz structure for which we choose the
canonical root system of $\mathfrak{su}(4)=\mathfrak{so}(6)$. We find that the
following choice, which we will refer to as the ``Beauty'', generates the full
superconformal algebra and has the desired properties.:
\begin{gather}\label{triplets-beauty}
	E_{I}^{+}=(\mathcal{L}_{2}^{1}\,,\mathcal{L}_{3}^{2}\,,\mathcal{L}_{4}
	^{3}\,,S_{1-}\,,J^{+})\,,\qquad E_{I}^{-}=(E_{I}^{+})^{\dagger}\,,\\
	H_{i}=\mathcal{L}_{i}^{i}-\mathcal{L}_{i+1}^{i+1}\,,\qquad i=1,2,3\,,\\
	H_{4}=\mathcal{L}_{1}^{1}-J^{(3)}\,,\qquad H_{5}=J^{(3)}\,.
\end{gather}
These triplets satisfy the Serre--Chevalley relations, and the Dynkin diagram
is depicted in Figure \ref{fig:beauty-6d-main-text}, with Cartan
matrix
\begin{equation}
	\mathcal{M}=
	\begin{pmatrix}
	2 & -1 & 0 & +1 & 0\\
	-1 & 2 & -1 & 0 & 0\\
	0 & -1 & 2 & 0 & 0\\
	-1 & 0 & 0 & 0 & -1\\
	0 & 0 & 0 & -2 & 2
	\end{pmatrix}\,.
\end{equation}
We note that the superconformal algebra, $\mathfrak{osp}(6,2|1)$, is a
non-compact form of the simple Lie superalgebra $\mathfrak{osp}(8|1)$, which is
compact. In our convention, this is reflected in the fact that the ladder
generators associated with special conformal transformations satisfy
$K^\dagger = -P$, and the off-diagonal positive sign appearing in the
Cartan matrix. At the level of representations of the algebra, this results in
the infinite tower of derivatives of the conformal primaries. This property is
apparent in the oscillator construction discussed in Appendix
\ref{app:oscillator-construction}.

Having derived the Cartan matrix, we can easily find its symmetrized version
used in the Bethe ansatz shown in equation \eqref{eqn:bethe2}:
\begin{equation}\label{beauty-symmetrised-cartan}
		\mathcal{A} = \Lambda \mathcal{M} = \begin{pmatrix}
			2& -1& 0& 1& 0 \\
			-1& 2& -1& 0& 0 \\
			0& -1& 2& 0& 0\\
			1& 0& 0& 0& 1\\
			0& 0& 0& 1& -1
	\end{pmatrix}\,,
\end{equation}
where $\Lambda = \text{diag}(1,1,1,-1,-1/2)$. The specific form of $\Lambda$
can be found by, for instance, calculating the form of the quadratic Casimir,
which references $\mathcal{A}$ explicitly.

\subsection{Representations}

Working with respect to the Beauty Dynkin diagram, we now review how to
construct the states at a given site of the super-spin chain. To do so, let us
start with a highest weight state $\left\vert \Omega\right\rangle$ that
transforms in a representation of $\mathfrak{u}(4|1)$. As described above, we
can label it in terms of its weight vector, or equivalently in terms of the
quantum number given in equation \eqref{quantum-numbers-6D}:
\begin{equation}\label{quantum-numbers-operators-6D}
\mathcal{D}\left\vert \Omega\right\rangle =\Delta\left\vert \Omega
\right\rangle \,,\qquad J^{(3)}\left\vert \Omega\right\rangle =j\left\vert
\Omega\right\rangle \,,\qquad
		(\mathcal{L}_{i}^{i} - \mathcal{L}_{i+1}^{i+1})\left\vert \Omega\right\rangle =\ell_{i}\left\vert
	\Omega\right\rangle \,,\qquad i=1,2,3 \,.
\end{equation}
There is no sum over the indices. Defined in this way, the triplet
$[\ell_{1},\ell_{2},\ell_{3}]$ corresponds to the Dynkin labels of
$\mathfrak{su}(4)$ representations in the usual convention (or equivalently of
$\mathfrak{so}(6)$), and coincide with the eigenvalues of $H_i,i=1,2,3$ of the
Beauty root system. As advertised in equation \eqref{u(4|1)-weights}, we can
always pass from the weights to the quantum numbers
$[\Delta;\ell_1,\ell_2,\ell_3;j]$ using equations
\eqref{quantum-numbers-operators-6D} and \eqref{triplets-beauty}:
\begin{align}\label{eqn:DELTAFORCE}
		\Delta &= \frac{1}{2}(-3 h_1 - 2 h_2 - h_3 + 4 h_4 + 4 h_5)\nonumber\,,\\
		\ell_i &= h_i\,,\qquad i=1,2,3\,,\\
		j &= h_5\nonumber\,,
\end{align}
where we used $\mathcal{D}=\frac{1}{2}\sum_{i=1}^4\mathcal{L}^i_i$. We can
therefore equivalently label a multiplet by the quantum numbers of its
superconformal primary or its weight vector.

The simplest example of a highest weight state of the Beauty is the vacuum in
the oscillator picture which, as explained in the previous section, is
associated with the field $X$:
\begin{equation}
		X\longleftrightarrow\left\vert 0\right\rangle \,,\qquad[\Delta;\ell_1,\ell_2,\ell_3; j] = [2;0,0,0;+1] \,.
\end{equation}
Using the generators in the oscillator picture described in Appendix
\ref{app:oscillator-construction}, we can check that $X\vert 0 \rangle$ is
annihilated by the $Q^{i+}$ and is therefore half-BPS. It is also annihilated by the
superconformal charges, $S_{i\pm}$, so it is a superconformal primary. We can also
further check that it has the correct quantum numbers. From there, the other
fields in the multiplet can be reached by applying the lowering operators,
$E^-_I$:
\begin{equation}
	Y^\dagger \longleftrightarrow J^{-}\left|0\right>\,,\qquad
	\psi^i \longleftrightarrow Q^{i+}J^-\left|0\right> \sim Q^{i-}\left|0\right> \,.
\end{equation}
Furthermore, we can construct infinite towers of covariant derivatives acting
on all these fields by successive applications of the translation operators,
$P^{ij}$:
\begin{equation}
	D^{ij}X\leftrightarrow P^{ij}\left\vert 0\right\rangle \,,\quad D^{kl}
	D^{ij}X\leftrightarrow P^{kl}P^{ij}\left\vert 0\right\rangle \,,\quad
	\cdots\quad D^{ij}Y^{\dagger}\leftrightarrow P^{ij}J^{-}\left\vert
0\right\rangle \,,\quad\cdots
\end{equation}
with $D^{ij}$ the associated covariant derivative.\footnote{We follow the
		conventions of the oscillator construction \cite{Gunaydin:1990ag,
		Gunaydin:1999ci, Gunaydin:1981yq}, where a bispinor notation is used,
		i.e. $K_{ij}\sim \Gamma_{ij}^\mu K_\mu\,,P^{ij}\sim
\widetilde{\Gamma}^{ij}_\mu P^\mu$.} Together, the states organize themselves
into the superconformal multiplet of type $D[j=1]$ (in the convention of
\cite{Cordova:2016emh}) associated with the free hypermultiplet.

\begin{figure}[ptb]
		\centering
		{\resizebox{0.14\textwidth}{!}{\includegraphics{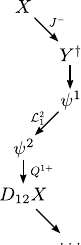}}}
		\caption{\label{fig:root_X} Weight diagram of the free hypermultiplet with
		respect to the subalgebra of $\mathfrak{osp}(6,2|1)$ generated by
		$E_I^+=\mathcal{L}^1_2,S_{1-}, J^+$. Each level is reached by applying
		the corresponding lowering operator, $E^-_I$. The other Lorentz
		generators can be added straightforwardly. Note that we ignore all
		overall prefactors for brevity.
		}
\end{figure}

The weight of each field is found by applying the Cartan generators $H_I$
defined in equation \eqref{triplets-beauty} and then using the commutation relations
that are summarized in Appendix \ref{app:oscillator-construction}. For
completeness, the spectrum of the free hypermultiplet is collected in Table
\ref{tab:6d-hyper}, and its weight diagram for the first few levels is shown in
Figure \ref{fig:root_X}. One can see that it is periodic, generating the
infinite tower of derivative for each field.

\begin{table}[ptb]
\centering
\begin{tabular}[c]{c|c|c|c}
	field & state & $[\Delta;\ell_1,\ell_2,\ell_3;j]$ & $w$ \\\hline
	$X$ & $\left|0\right>$ & $[1; 0,0; 1,0]$ & $(0,0,0,0,+1)$\\
	$Y^{\dagger}$ & $J^-\left|0\right>$ & $[2;0,0,0;-1]$ & $(0,0,0,2,-1)$\\
	$\psi^{1}$ & $Q^{1-}\left|0\right>$ & $[\frac{5}{2};1,0,0;0]$& $(1,0,0,2,0)$\\
	$D^{\ell}X$ & $(P^{12})^\ell\left|0\right>$ & $[2+\ell;0,\ell,0;+1]$& $(0,\ell,0,\ell, +1)$
\end{tabular}
\caption{\label{tab:6d-hyper} Field content of the free field hypermultiplet
and its representation data. Only Lorentz highest weight states are shown,
and $w_I$ is given with respect to the Beauty Dynkin diagram, see equation
\eqref{triplets-beauty}.}
\end{table}

Note that, in principle, we should also introduce the generators of the flavor
and gauge algebras, as the hypermultiplets, $X_i$, transform in the
bifundamental representation of $G_{i-1}\times G_{i}$. However, since we are
only interested in gauge invariant operators made out of $X_i$ or its
descendants, their position in the spin chain is fixed. Had we instead been
interested in more general gauge invariant operators and allowed for the
presence of, e.g., $X^\dagger_i$ transforming in the conjugate representations
of the gauge groups, operators could involve gauge singlets $(X_iX^\dagger_i)$
as building blocks, and the structure of the super-spin chain would have been
more involved. We leave such an analysis for future work.

\section{Integrability for Long Quivers}\label{sec:INTEGRO}

In this section we turn to the study of operator mixing as described by a
super-spin chain with nearest neighbor interactions. For ease of exposition, we
again focus on the case of A-type conformal matter (i.e., hypermultiplets),
deferring generalizations to Section \ref{sec:GEN}.

Recall that in the previous section we discussed the excitations on a single
spin chain site. In the context of a 6D SCFT on its partial tensor branch,
each of these sites specifies conformal matter in a bifundamental
representation of neighboring gauge groups, $G_{i-1}\times G_i$. Introducing
the corresponding spin chain sites:
\begin{equation}
	X_{i}\longleftrightarrow\left\vert 0\right\rangle_{i}\,,
\end{equation}
we can now form a gauge invariant operator in the bifundamental representation
of the left and right flavor symmetries via the product $X^L=X_{1}\cdots X_{L}$,
which we view as the ground state of a spin chain with $L$ sites:
\begin{equation}
X^{L}\longleftrightarrow\left\vert \text{GND}\right\rangle =\left\vert
0\right\rangle _{1}\otimes\left\vert 0\right\rangle _{2}\otimes\cdots \otimes
\left\vert 0\right\rangle _{L-1}\otimes\left\vert 0\right\rangle _{L} \,.
\end{equation}
As far as the superconformal algebra is concerned, we can of course consider
the action of $\mathfrak{osp}(6,2|1)$ on the entire state
$\left|\text{GND}\right>$. In the spin chain picture, we consider multiple
copies of the superconformal algebra (one for each site). Gauge invariant
states are then obtained by projecting from the larger Hilbert space in the
obvious way.

With this in mind, we can now construct a wide variety of candidate gauge
invariant operators in the tensor branch theory. Each of these states can also
be identified in the corresponding holographic dual as an excitation along the
great arc stretching from the north to the south pole orbifold fixed points of
the geometry $\text{AdS}_{7}\times S^{4}/\mathbb{Z}_{K}$. As such, each of these
operators is expected to persist at the conformal fixed point and we can ask
about operator mixing of the corresponding states just constructed.

The operators of interest are then realized as excitations above the protected
ground state
\begin{equation}\label{operator-basis}
	\mathcal{O}\longleftrightarrow\prod_{i=1}^{L}\prod_{I=1}^{r=5}[E_{I}^{-}(i)]^{n_{i}^{I}}\left|\text{GND}\right>\,,
\end{equation}
where $E^{-}(i)$ are the lowering operators at the $i$th site, and
$n^I=\sum_{i=1}^L n^I_i$ the total number of excitations of the $I$th simple
root.

Operator mixing in the $\mathfrak{su}(2)$ R-symmetry subsector was studied in
references \cite{Baume:2020ure,Heckman:2020otd} and was also reviewed in
Section \ref{sec:REVIEW}. The goal of this section will be to show that the
integrability found in this R-symmetry sector, where we allow only for
excitation of $J^-$, actually extends to the full $\mathfrak{osp}(6,2|1)$
super-spin chain. On general grounds, we expect that operator mixing is
controlled by a spin-chain Hamiltonian with nearest neighbor interactions. This
follows from locality in the holographic dual for excitations along the great
arc of $S^{4}/\mathbb{Z}_{K}$. In other words, at leading order in a $1/L$
expansion, the dilatation operator/spin-chain Hamiltonian takes the form:
\begin{equation}\label{hamiltonian-per-site}
	\delta\mathcal{D} = \lambda_{G}\sum_{i=1}^{L-1}\mathcal{H}_{i,i+1}\,,
\end{equation}
where $\lambda_{G}$ is a calculable number (see Section \ref{sec:REVIEW}). Due
to the decomposition given in terms of nearest-neighbor interactions as given
in equation (\ref{hamiltonian-per-site}), we can focus on the action of the
spin-chain Hamiltonian on chains of length $L=2$.

By symmetry, operator mixing cannot change the total quantum numbers of any
states, but can certainly rearrange them into different representations.
Therefore, instead of using the basis of operators given in equation
\eqref{operator-basis}, we may choose a basis of irreducible representations of
$\mathfrak{osp}(6,2|1)$, such that the Hamiltonian act by multiplication and
gives us the anomalous dimension directly. In other words, we can further
decompose the $L=2$ Hamiltonian as:
\begin{equation}\label{hamiltonian-decomposition-projector}
	\mathcal{H}_{i,i+1}=\sum_{\mathcal{R}\in D[1]\otimes D[1]}f(w_\mathcal{R})P_\mathcal{R}\,,
\end{equation}
where $P_\mathcal{R}$ is the projection from the tensor product of two
hypermultiplets $D[1]$ onto the irreducible representation $\mathcal{R}$.
Since we are working with irreducible representations of the Beauty, the
function $f$ depends only on the data of the highest weight vector,
$w_\mathcal{R}$.

The Hamiltonian commutes with the other generators and it is in fact a function
of the Casimir elements of the superconformal algebra in the free field limit,
and $f$ must be a function of their eigenvalues. In the context of integrability, it
is enough to show that if $f = -2h(c)$, where $c$ is obtained from the
eigenvalue of the quadratic Casimir and $h(n)=\sum_{k=1}^{n}\frac{1}{k}$ the
harmonic number, then the Hamiltonian given in equation
\eqref{hamiltonian-decomposition-projector} satisfies the Yang--Baxter
equations and is integrable. This is a well-known result for the XXX spin chain
and its deformations \cite{Faddeev:1996iy}, and has also been studied in the
context of SCFTs starting with the seminal work of references
\cite{Beisert:2003jj, Beisert:2003yb}.

Let us illustrate these features in the case of the $\mathfrak{su}(2)_{R}$
subsector. For $\mathfrak{g}=\mathfrak{su}(2)$, the quadratic Casimir and its
eigenvalues take the usual form\footnote{Once again, we remind the reader that
we are using a convention for which Cartan eigenvalues are integer-valued.}
\begin{gather}
	C_{\mathfrak{su}(2)}=\frac{1}{2}(J^{(3)})^{2}+\left\{J^{+},J^{-}\right\}\,,\\
	C_{\mathfrak{su}(2)}\left\vert w=j\right\rangle =\frac{1}{2}j(j+2)\left| j\right> \,.
\end{gather}
With $L=2$, the tensor product of the two doublets formed by the scalar fields
of the hypermultiplet decomposes into the adjoint and singlet
representations of $\mathfrak{su}(2)_{R}$ with highest weight
states:\footnote{For clarity, we will alway suppress the position indices when
discussing $L=2$ states.}
\begin{equation}\label{su2R-eigenstates}
\begin{split}
	XX\quad\longleftrightarrow\quad & \left\vert j=2\right\rangle =\left|\text{GND}\right>\,, \\
	XY^{\dagger}-Y^{\dagger}X\quad\longleftrightarrow\quad & \left|j=0\right> =\left( J^{-}(i+1)-J^{-}(i)\right) \left|\text{GND}\right>\,.
\end{split}
\end{equation}
The Hamiltonian encoding the operator mixing in the R-symmetry sector is given
by \cite{Baume:2020ure}:
\begin{equation}\label{Ham-Casimir-su2R}
		\mathcal{H}_{i,i+1}= -2C_{\mathfrak{su}(2)_R} - 1 = -2\left(h(0)P_{j=2} + h(1)P_{j=0}\right) = -2P_{j=0}\,,
\end{equation}
has the form advertised in equation \eqref{hamiltonian-decomposition-projector}
with $f(w)=-2h(\frac{2-j}{2})$, and is therefore integrable, as expected.

Our aim in the rest of this section will be to lift this analysis to the full
super-spin chain, and to work out some of the implications of integrability in
this system. To that end, we begin by showing how the Hamiltonian of the
$\mathfrak{su}(2)$ R-symmetry sector can be used to determine another subsector
obtained from acting by lightcone derivatives on the $X$'s, the so-called
non-compact $\mathfrak{sl}(2)$ sector. After establishing integrability in this
subsector we show (using similar reasoning that of references \cite{Beisert:2003jj,
Beisert:2003yb}) that integrability of the full super-spin chain follows. A
helpful feature of this analysis is that the only Feynman diagrams we need to
evaluate in this process involve the $\mathfrak{su}(2)$ R-symmetry sector, a
task which was already completed in \cite{Baume:2020ure}. With this in place,
we can leverage well-known results on integrability to obtain the Bethe ansatz
for the corresponding super-spin chain with open boundary conditions. The case
of little string theories amounts to simply changing to periodic boundary
conditions.

As a final comment, we note that similar considerations hold for the related
case of 4D $\mathcal{N}=2$ SCFTs with the same quiver structure. In fact,
integrability for this case was investigated in \cite{Beisert:2005he}, where it
was found that the leading order mixing terms of long quivers can be obtained
from suitable orbifold projections of $\mathcal{N}=4$ super Yang--Mills theory.
Integrability at leading order in a large R-charge expansion is then inherited
from $\mathcal{N} = 4$ super Yang--Mills theory (see e.g., references
\cite{Berenstein:2002jq, Minahan:2002ve, Beisert:2003jj, Beisert:2003yb}).

\subsection{An \texorpdfstring{$\mathfrak{sl}(2)$}{sl(2)} Subsector and its Lift to \texorpdfstring{$\mathfrak{osp}(6,2|1)$}{osp(6,2|1)}}

\begin{figure}[t!]
\centering
\begin{subfigure}[t]{.49\linewidth}
		\centering
		{\resizebox{0.64\textwidth}{!}{\includegraphics{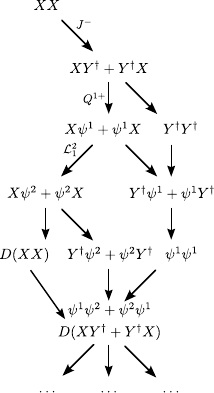}}}
	\end{subfigure}
\hfill\begin{subfigure}[t]{.49\linewidth}
		\centering
		{\resizebox{0.64\textwidth}{!}{\includegraphics{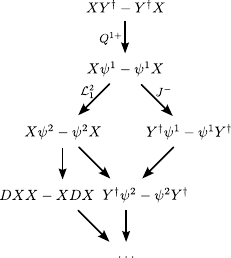}}}
	\end{subfigure}
\caption{
		\label{fig:weight-diagrams} Weight diagram of irreducible Lorentz
		scalar representations of
		$\mathfrak{g}_\text{sub}=\left<\mathcal{L}^1_2,S_{1-},J^+\right>$
		obtained by taking the tensor product of two hypermultiplets. Each
		state is reached by applying the lowering operators, $E_I^-$, and the
		derivatives $D\leftrightarrow P^{12}$ are taken along the $[0,1,0]$
		$\mathfrak{su}(4)$ direction.
}
\end{figure}

To establish integrability for the full super-spin chain, it is instructive to
construct a few of the irreducible representations appearing in the tensor
product of two hypermultiplets. It turns out that there are only two
superconformal primaries that are Lorentz scalars: $XX$ and $XY^\dagger -
Y^\dagger X$. The first few levels of their respective weight diagrams are
depicted in Figure \ref{fig:weight-diagrams}.

An immediate consequence is that all possible combinations of the three types
of fields $X,Y^\dagger, \psi^1$ appear in either multiplet. By superconformal
symmetry, this means that they all share the same anomalous dimensions and,
since the superconformal primaries are the two $\mathfrak{su}(2)_R$
eigenstates, see equation \eqref{su2R-eigenstates}, we can immediately extend
integrability to a larger class of operators.

However, this is not sufficient to show integrability of complete system. For
instance, with fermions of different types there can be non-trivial
interactions where two fermions are ``merged'' into a covariant derivative,
schematically through an interaction of the form $\psi\psi\to XDX$, and we need
to consider operator mixing involving multiple derivatives. To do so, it is
sufficient to focus on a sector where all derivatives point in the same
lightcone direction, and we denote such an impurity at the $i^{\text{th}}$ spin
chain site as: $DX_{i}$. In contrast to the case of the
$\mathfrak{su}(2)_R$ sector, we can consider an arbitrary number of
descendants, e.g., $D^{\ell}X_{i}$. Indeed, in the oscillator representation
of the superconformal algebra (see Appendix \ref{app:6Doscillators}), these
states are associated with bosonic excitations.

To begin, we observe that this sector is associated with a non-compact algebra
isomorphic to $\mathfrak{sl}(2,\mathbb{R})$ generated by $K_{12}$, $P^{12}$ and their
commutator:
\begin{equation}\label{sl2-generators}
		\begin{gathered}
		[K_{12}, -P^{12}]=H=-\mathcal{D}-\frac{1}{2}(H_1+H_2+H_3)\,,\\
		[H,K_{12}]=2K_{12}\,,\qquad[ H,P^{12}]=-2P^{12}\,,
		\end{gathered}
\end{equation}
with $H_i$ the Lorentz Cartan generators, see Section \ref{sec:beauty-6D}. A
single derivative of $X$ specifies a highest weight state $[0,1,0]$ of
$\mathfrak{so}(6)$, so these states specify single site weights:
\begin{equation}
D^{\ell}X_i\longleftrightarrow(P^{12})^{\ell} \left\vert 0\right\rangle
\,,\qquad  [\Delta;\ell_1, \ell_2, \ell_3; j]=[2+\ell;0,\ell,0;+1]\,.
\end{equation}
The quadratic Casimir takes the same form as that of $\mathfrak{su}(2)_R$ and
acts, in the free field limit, on an $L=2$ highest weight state with $\Delta=4+\ell$ as:
\begin{equation}\label{sl2-Casimir}
		C_{\mathfrak{sl}(2)} \left|\Omega\right> = \frac{1}{2}H(H+2)\left|\Omega\right> =  2(\ell+1)(\ell+2)\left|\Omega\right>\,,
\end{equation}
such that representations can be labeled by the total number $\ell$ of
derivatives of their primaries.

We would like to understand operator mixing for such derivative insertions.
Now, both the $\mathfrak{sl}(2)$ and $\mathfrak{su}(2)_{R}$ sectors can of
course be embedded in the 6D superconformal algebra. As an intermediate step,
consider the superalgebra $\mathfrak{g}_{\text{sub}}$ generated
by:\footnote{Comparing with the case of 4D $\mathcal{N}=4$ super Yang--Mills
		theory or the 4D $\mathcal{N}=2$ quivers discussed below, the same line
		of argument applies and the corresponding subalgebra is generated by
$\left<S_{1-},J^{+},\bar{S}_{\bar{1}-}\right>$, which itself contains the 4D
$\mathfrak{sl}(2)$ subsector generated by $K_{1\bar{1}}$, $P^{11}$ and their
commutator.}
\begin{equation}\label{generators-sub-osp}
	\mathfrak{g}_{\text{sub}}=\left< \mathcal{L}_{2}^{1}\,, S_{1-}\,, J^{+}\,\right> \,,
\end{equation}
which we recognize as a subset of the Beauty diagram (see Figure
\ref{fig:beauty-6d-main-text}). Using the commutation relation, one can see that it
contains the $\mathfrak{sl}(2)$ sector since $\{S_{1}^{-},[J^{+},[\mathcal{L}
_{2}^{1},S_{1}^{-}]]\}\sim K_{12}$.

Let us now show that integrability of the non-compact derivative sector follows
from that of the $\mathfrak{su}(2)_R$ subsector. At the one-derivative level, the
two $\mathfrak{sl}(2)$ eigenstates are $(DX)X\pm X(DX)$, but as shown in Figure
\ref{fig:weight-diagrams}, they are merely superconformal descendants of the
$\mathfrak{su}(2)_R$ highest weights, see equation \eqref{su2R-eigenstates}.

This is already important because it means that generically, the motion of a
single derivative insertion again obeys a free propagating wave equation,
precisely as one would expect based on the holographic picture. To establish
the Hamiltonian, however, we also need to consider the scattering of
impurities on the spin chain.

Given a fixed number of derivative insertions, there is always a unique
combination that is a highest weight state of $\mathfrak{sl}(2)$, all the
others eigenstates being $P^{12}$-descendants. For instance, with two impurities the
$\mathfrak{sl}(2)$ eigenstates are given by:
\begin{gather}
	D^{2}(XX)\,,\qquad D((DX)X-X(DX))\,,\\
	(D^{2}X)X-3(DX)(DX)+X(D^{2}X)\,.
\end{gather}
The first two are in the  $\ell=0,1$ multiplets for which we know the anomalous
dimension, while the third is the $\ell=2$ highest weight state. While a
conformal primary, the latter can be written as
$D^2(XX)+\frac{5}{2}Q^{1+}Q^{2+}Q^{1-}Q^{2-}(XX)$, which we
recognize as a combination of $\mathfrak{g}_\text{sub}$-descendants of $XX$,
see Figure \ref{fig:weight-diagrams}. The $\ell=2, j=2$ sector is thus fully
encoded in the $\ell=0,1$ and $\mathfrak{su}(2)_{R}$ sectors.
A similar pattern also occurs for any number of derivatives: among
the $\mathfrak{sl}(2)$ eigenstates, there is always a unique $\mathfrak{sl}(2)$
highest weight state, the others being $P^{12}$-descendants. This can be seen straightforwardly by showing that there is a single combination of states (up to normalization) with $\ell$ derivatives distributed among the two sites that is annihilated by $K^{12}$. Together with the $P^{12}$-descendants, they form a basis for any fixed-$\ell$ states. One can then always find a special
state that can be written as a $\mathfrak{g}_\text{sub}$-descendant of $XX$:
$D^{\ell-2}((DX)(DX))=-\frac{1}{2}D^{\ell-2}Q^{1+}Q^{2+}Q^{1-}Q^{2-}(XX)$.
Therefore the $\mathfrak{sl}(2)$ highest weight state with $\ell$ derivatives
can always be written as a linear combination of $P^{12}$-descendants and this
special state. By induction, if we know the operator mixing for $\ell'<\ell$
derivatives, we can deduce it for $\ell$ derivatives. The structure of the
$\mathfrak{sl}(2)$ spin chain is therefore uniquely fixed by the
$\mathfrak{su}(2)_R$ sector, and thus we conclude that the lightcone derivative sector
is controlled by an integrable, non-compact XXX$_{s=\frac{w}{2}}$ spin chain,
with $w=-\Delta_X=-2$, and a Hamiltonian given by \cite{Faddeev:1996iy}:
\begin{equation}\label{sl2-hamiltonian}
		\mathcal{H}_{i,i+1} = \sum_{\ell\geq0} (h(\ell+1) -1 ) P_{\ell}\,,
\end{equation}
where $P_\ell$ is the projector onto the spin-$\ell$ $\mathfrak{sl}(2)$
eigenstate.

To lift integrability of the $\mathfrak{sl}(2)$ and $\mathfrak{su}(2)_R$ sector
to $\mathfrak{osp}(6,2|1)$, let us note that studying operator mixing of the
$L=2$ chain is equivalent to considering the correlator of four
hypermultiplets. In theories with eight supercharges, the structure of such
four-point functions is known to be very constrained, see e.g., references
\cite{Chang:2017xmr, Ferrara:2001uj, Dolan:2004mu, Bobev:2017jhk,
Baume:2019aid}. In our case, taking into account that the flavor and gauge
structure further constraints the type of multiplets that may appear, we have
the following schematic decomposition:
\begin{equation}
		D[1]\otimes D[1] = D[2] \oplus \sum_{\ell'\geq0}A[0,\ell',0]^{j=0}\,,
\end{equation}
where $D[2]$ is the multiplet associated with $XX$, while the
$A$-type multiplets are those that recombine into long multiplets
when interactions are turned on. Note that we need only consider long
multiplets whose superconformal primaries have $j=0$. This can be seen from
Figure \ref{fig:weight-diagrams}: states with $j=2$ must either be descendants
of $XX$ or states with $(\ell'=\ell-2,j=0)$, while any state with $j=1$ must
take the form $D^mY^\dagger D^n\psi^i \sim Q^{i+}(D^mY^\dagger D^nX)$ and can
therefore be written as a superconformal descendant of $j=0,2$ primaries.

Now, the eigenvalues of the quadratic superconformal Casimir with respect to
the root system of the ``Beauty'' are given by
\begin{equation}
	C_{\mathfrak{osp}(6,2|1)}\left|w\right>=\left\langle \mu,\mu+2\rho\right\rangle\left|w\right>
\,,\qquad\rho=(1,1,1,0,1)\,,
\end{equation}
with $\rho$ the Weyl vector and $\left<\alpha,\beta\right>
=\alpha_{I}\mathcal{A}^{IJ}\beta_{J}$ the root pairing for our choice of root
system, see the discussion around equation \eqref{Serre--Chevalley}. This
expression follows from the $\mathfrak{osp}(6,2|1)$ quadratic Casimir in the
oscillator picture, see equation \eqref{quadratic-casimir-osp}, and we recover
(up to minor changes in conventions) the well-known expression for multiplets
in $\ell$-traceless-symmetric Lorentz representations for superconformal
theories with eight supercharges, see e.g., references \cite{Bobev:2017jhk,
Baume:2019aid}:
\begin{align}
		C_{\mathfrak{osp}(6,2|1)}=c_{\Delta,\ell} + 4\Delta - j(j+2)\,,\\
		c_{\Delta,\ell} = \Delta(\Delta-6) + \ell(\ell+4)\,,
\end{align}
with $c_{\Delta,\ell}$ the eigenvalue of the $\mathfrak{so}(6,2)$ conformal
Casimir. For a superconformal multiplet of type $A[0,\ell',0]^{j=0}$ with $\ell'=\ell-2$, unitary bounds
imply that $\Delta=4+\ell$, and we obtain, for any state $\vert \phi \rangle$ in the $A$-multiplet:
\begin{align}
		C_{\mathfrak{osp}(6,2|1)}\left| \phi \right> = 2 (\ell+1)(\ell+2) \left| \phi \right>.
\end{align}
This is precisely the same value as that of the $\mathfrak{sl}(2)$ Casimir, see
equation \eqref{sl2-Casimir}, and shows that we can embed every
$\mathfrak{sl}(2)$ highest weight states into the appropriate superconformal
multiplet $A[0,\ell-2,0]^{j=0}$, and the Hamiltonian must be (up to a
possible shift defining the ground state energy) of the form given in equation
\eqref{sl2-hamiltonian}.

We therefore infer that the operator mixing for long 6D quivers can be extended
from the R-symmetry sector to the rest of the superconformal symmetry. Much as in the original derivation of integrability in $\mathcal{N}=4$ super-Yang--Mills theory, which used the derivative sector to extend integrability to the complete superconformal algebra \cite{Beisert:2003jj}, here we used that since the extension of $\mathfrak{su}(2)_R$ to $\mathfrak{osp}(6,2|2)$ via $\mathfrak{sl}(2)$ is unique, the one-loop anomalous dimension of a given superconformal multiplet is fixed by that of the appropriate derivative state. Since the $\mathfrak{sl}(2)$ and $\mathfrak{osp}(6,2|1)$ Casimir eigenvalues are the same up to an appropriate shift, the harmonic properties of the Hamiltonian in the derivative sector therefore apply to any state, which establishes integrability across the entire super-spin chain.

\subsection{Bethe Ansatz}

Having established integrability of the super-spin chain, we now put this to
use and extract the corresponding energy eigenstates, i.e., operator scaling
dimensions of the 6D SCFT. Thankfully, this task has already been carried out
in great detail in \cite{Arnaudon:2003zw} (see also \cite{Frahm:2022hof}).
Putting these results in a form closer to those which appear in references
\cite{Minahan:2002ve, Beisert:2003yb}, we consider excitations of the spin
chain pointing in a particular direction of a representation, given by
excitations of the simple root of the Beauty.

To begin, we introduce a highest weight vector $w_{I}$ for our spin chain. We
view each excitation as a specifying a direction in the representation, and
correspondingly introduce impurities associated with the simple roots, with
corresponding rapidities $u_{i}^{I}$.  Here, $i$ indicates the impurity in
question and $I=1,\cdots ,r$ indicates a root of our Lie superalgebra ($r=5$ for
$\mathfrak{osp}(6,2|1)$). The Bethe ansatz also makes reference to $\mathcal{A}$, the
symmetrized Cartan matrix of the Beauty as defined in equation
\eqref{beauty-symmetrised-cartan}.

The Bethe ansatz equations for the super-spin chain with open boundary
conditions and a total of $n^J$ impurities of each type are then:
\begin{equation}\label{osp-Bethe-Ansatz-open}
		\left(\frac{u_{j,J}+\frac{i}{2}w_{J}}{u_{j,J}-\frac{i}{2}w_{J}}\right)^{2L}=
		\prod_{K=1}^{r}\prod_{\substack{k=1\\(k,K)\neq(j,J)}}^{n^{K}}
		\frac{u_{j,J}-u_{k,K}+\frac{i}{2}\mathcal{A}_{JK}}{u_{j,J}-u_{k,K}-\frac{i}{2}\mathcal{A}_{JK}}\,
		\frac{u_{j,J}+u_{k,K}+\frac{i}{2}\mathcal{A}_{JK}}{u_{j,J}+u_{k,K}-\frac{i}{2}
\mathcal{A}_{IJ}}
\,.
\end{equation}
There is an overall momentum constraint on the rapidities, as required in order
for the momentum on the spin-chain to not leak off the ends of the M5-brane.
This imposes the decoupling constraint on the $J = 5$ direction (i.e., the
R-symmetry excitations) of the weight space:
\begin{equation}
	\prod_j\left(
		\frac{u_{j,J=5}+\frac{i}{2}w_{J = 5}}{u_{j,J=5}-\frac{i}{2}w_{J = 5}}
	\right)=1\,.
\end{equation}
Note that this is just a constraint on the first descendants of $X$. All other
excitations are generated by further descent, which in turn imposes a net
momentum conservation condition across the full chain. See e.g., reference
\cite{Minahan:2002ve} for some further discussion in the case of an $SO(6)$
spin chain. We note that in the closely related super-spin chain for the 4D
$\mathcal{N}=2$ SCFT obtained from reduction on a $T^{2}$, there is no momentum
conservation constraint, since momentum can leak out in the extra-dimensional
$T^{2}$ direction anyway.

With this in place, we can finally extract the operator scaling dimensions for
any super-spin chain excitation. The energies/anomalous dimensions are given
by:
\begin{equation}
	\Delta= E = \Delta_{G}^{0}+\lambda_{G} \left\vert \sum_{J=1}^{r}\sum_{j=1}^{n^{J}}
	\frac{i}{u_{j,J}+\frac{i}{2}w_{J}}-\frac{i}{u_{j,J} - \frac{i}{2}w_{J}} \right\vert \,,
\end{equation}
where here, we have indicated by $\Delta_{G}^{0}$ the scaling dimension of the
half-BPS operator $X^{L}$ for $(G,G)$ conformal matter, and $\lambda_{G}$ is
a calculable coefficient obtained from evaluation in the $\mathfrak{su}(2)_{R}$
sector, see equation \eqref{lambda_G}.

\subsection{LST Boundary Conditions}

Similar considerations hold for little string theories (LSTs). Indeed, the
only difference is that for the 6D SCFT with quiver:
\begin{equation}
\lbrack G_{0}]-G_{1}-\cdots -G_{L-1}-[G_{L}] \,,
\end{equation}
we now gauge the diagonal subgroup of $G_{0}\times G_{L}$
(accompanied by a corresponding tensor multiplet). As far as operator mixing
is concerned, the nearest neighbor hopping terms are the same, but we now have
periodic boundary conditions. The corresponding Bethe ansatz equations are (notation as in
equation \eqref{osp-Bethe-Ansatz-open}):
\begin{equation}
		\left(\frac{u_{j,J}+\frac{i}{2}w_{J}}{u_{j,J}-\frac{i}{2}w_{J}}\right)^{2L}
		=\prod_{K=1}^{r}\prod_{\substack{k=1\\(k,K)\neq(j,J)}}^{n^{K}}
		\frac{u_{j,J}-u_{k,K}+\frac{i}{2}\mathcal{A}_{JK}}{u_{j,J}-u_{j,J}-\frac{i}{2}\mathcal{A}_{JK}}\,,
\end{equation}
and we again impose an overall momentum constraint on the R-symmetry fluctuations (which also constrains
the descendant states):
\begin{equation}
	\prod_j\left(
			\frac{u_{j,J = 5}+\frac{i}{2}w_{J = 5}}{u_{j,J = 5}-\frac{i}{2}w_{J = 5}}
	\right) = 1 \,.
\end{equation}

\section{Generalizations}\label{sec:GEN}

Our discussion so far has centered on the case of 6D SCFTs with A-type
conformal matter, namely where the tensor branch gauge theory has hypermultiplets in bifundamental representations. In this section, we explore various
generalizations to other closely related systems. One generalization involves
the analogous computation of operator mixing for quivers with D- and E-type
conformal matter, as obtained from the worldvolume theory of $L$ M5-branes
probing an ADE singularity $\mathbb{C}^2 / \Gamma_{ADE}$, with $\Gamma_{ADE}$ a
finite subgroup of $SU(2)$. The partial tensor branch of these theories is a
generalized quiver gauge theory with D- and E-type gauge groups where the links
are given by bifundamental conformal matter.  Further compactification on a $T^2$ also results in 4D $\mathcal{N} = 2$ SCFTs, with operator mixing which closely tracks with the 6D case.

\subsection{D- and E-type Conformal Matter}

We begin by considering the generalization to D- and E-type conformal matter.
In the large $L$ limit, the worldvolume theory of $L$ M5-branes probing an ADE
singularity $\mathbb{C}^2 / \Gamma_{ADE}$ results in a holographic dual
$\text{AdS}_7\times S^4/\Gamma_\text{ADE}$ with $L$ units of four-form flux threading
the $S^4$.  Going slightly on the tensor branch corresponds to deconstructing a
great arc going from the north to south pole, where each segment is associated
to a site of the spin chain.  From this point of view, the different choices of
an ADE subgroup $\Gamma_{ADE}$ are all on the same footing, and as such, the
appearance of integrability in the A-type case is expected to also hold (from
holographic considerations) in the D- and E-type cases as well, as discussed in \cite{Baume:2020ure}. Our operating
assumption in this section will be that this is indeed the case.

Now, when a spin chain associated with a Lie (super)algebra is integrable, the
(nearest-neighbor) Hamiltonian, and by extension the Bethe ansatz ultimately depends solely on the
representation, $\mathcal{R}$ of each site and must act harmonically on
irreducible representations of $\mathcal{R}\otimes\mathcal{R}$. Minimal $(G, G)$ conformal matter is a generalization of the free $SU(N) \times SU(N)$ bifundamental hypermultiplet to a $G$ which is not special-unitary; when $G$ is of D- or E-type, this is an interacting SCFT \cite{DelZotto:2014hpa}. Such SCFTs contain a specific half-BPS operator known as the conformal matter operator, $X$, which is simply the scalar inside of the free hypermultiplet when $G$ is of A-type. As this conformal matter operator is protected against quantum corrections, its conformal dimension is set
by its R-charge, $j$. The scaling dimension was obtained in reference
\cite{Heckman:2014qba} (see also \cite{Baume:2020ure}), and the result is:
\begin{equation}\label{Delta_CM}
		\Delta^{6D}_\text{CM} = 2 j \,.
\end{equation}
The value of $j$ for each case has already been discussed in Section \ref{sec:INTRO},
but is repeated in Table \ref{tab:conformal-matter-operator} for convenience. For A-type quivers, this operator is precisely the field $X$ we
have discussed at length in the previous sections. More generally, $X$ is the primary of a superconformal multiplet of type $D[j]$;
its quantum numbers and weights, the latter with respect to the Beauty, are given by:
\begin{equation}\label{ADE-weights}
		[\Delta,\ell_1,\ell_2,\ell_3, j] = [2j;0,0,0;j]\,,\qquad w = (0,0,0,0,j) \,.
\end{equation}

\begin{table}
		\centering
  \begin{threeparttable}
		\begin{tabular}[c]{c|ccccc}
    \toprule
			$G$ & $SU(K)$ & $SO(2K)$ & $E_{6}$ & $E_{7}$ & $E_{8}$\\\midrule
			$j(X)$ & $1$ & $2$ & $3$ & $4$ & $6$\\\bottomrule
		\end{tabular}
  \end{threeparttable}
		\caption{The R-charge, $j(X)$, for the conformal matter operator $X$ of minimal $(G, G)$ conformal matter.}
		\label{tab:conformal-matter-operator}
\end{table}

As the Beauty is chosen such that highest weight states and superconformal primaries coincide, we can again define a half-BPS ground state of the spin chain
\begin{equation}
		X^L \leftrightarrow \left|\text{GND}\right> \,,
\end{equation}
where now $X$ specifies 6D conformal matter operator. Excitations above this ground
state are once again reached using the lowering operators $E_I^-$.  Note that
in the case of D- and E-type conformal matter, the $\mathfrak{su}(2)_R$ closed
subsector involves spins which are no longer in doublet representations. As
such, there is already a more intricate operator mixing structure in all of
these cases \cite{Baume:2020ure}.

Following through the same steps used in Section \ref{sec:INTEGRO}, we observe
that integrability in the $\mathfrak{su}(2)_R$ sector has important
implications for operator mixing in the full $\mathfrak{osp}(6,2|1)$ super-spin
chain. Indeed, there is materially little difference in extending first to the
$\mathfrak{sl}(2)$ subsector of D- and E-type quivers, and from this, lifting
all the way to the full super-spin chain. In particular, the corresponding
Bethe ansatz equations are essentially unchanged, up to the fact that we now
work with a highest weight state with highest R-charge $j$ instead of just $j = 1$ in the A-type case, as specified by equation \eqref{ADE-weights}.

\subsection{4D \texorpdfstring{$\mathcal{N} = 2$}{N=2} SCFTs and \texorpdfstring{$\mathfrak{su}(2,2|2)$}{su(2,2|2)} Super-Spin Chains}\label{sec:generalizations-4d}

Starting from the partial tensor branch of a quiver-like 6D SCFT,
compactification on a $T^2$ yields a 4D $\mathcal{N} = 2$ SCFT with the same
quiver-like description. Compared with their 6D counterparts, the 4D case has
marginal couplings, so at least in the case of A-type gauge groups, one can
directly verify the structure of nearest neighbor hopping terms for the related
class of spin chain operators. Indeed, in the A-type case, integrability is
inherited from a suitable orbifold projection of $\mathcal{N} = 4$
super Yang--Mills theory \cite{Beisert:2005he}. For the generalized quivers
with D- and E-type gauge groups and corresponding 4D conformal matter, we again
must rely on our stringy characterization, but we again expect the same
integrable structure to persist in the R-symmetry sector\cite{Baume:2020ure}.
A quite similar analysis to the one given in Section \ref{sec:INTEGRO} then implies that
integrability extends to the full super-spin chain.

The representation theory associated with the spin chain, however, does change.
The superconformal algebra is now $\mathfrak{su}(2,2|2)$, and we therefore must
again choose a basis for the root system which will define the minimal
excitations above the ground state associated with $X^L$. We hence face the
same problem as before: the usual quantum numbers defined by the bosonic
subalgebra
\begin{equation}
		\mathfrak{so}(2)\oplus\mathfrak{so}(4)\oplus\mathfrak{su}(2)_R\oplus\mathfrak{u}(1)_r:\qquad [\Delta;\ell,\bar{\ell};j,r]\,,
\end{equation}
do not prove to be a convenient basis to discuss $\mathfrak{su}(2,2|2)$
excitations of the spin chain. Instead, the maximal compact subalgebra
$\mathfrak{su}(2|1)\oplus\mathfrak{u}(1)_R\oplus\mathfrak{su}(2|1)$, will again
identify highest weight states with superconformal primaries:
\begin{equation}\label{4d-beauty-generators}
		\mathfrak{su}(2|1)\oplus\mathfrak{u}(1)_R\oplus\mathfrak{su}(2|1): \quad\mathcal{L}^i_j\oplus S_{i\pm}\oplus J^{(3)}\oplus \bar{\mathcal{L}}^{\bar{k}}_{\bar{l}}\oplus \bar{S}_{\bar{k}\pm}\,,
\end{equation}
where $i,j,\bar{k},\bar{l} = 1,2$ are the two Lorentz spinor indices, and $J^{(3)}$ generates
the Cartan of $\mathfrak{su}(2)_R$. This maximal compact subalgebra can also be
represented in terms of creation/annihilation operators, which we give in
Appendix \ref{app:oscillator-construction}.

\begin{figure}[t!]
	\centering
	{\resizebox{0.35\textwidth}{!}{\includegraphics{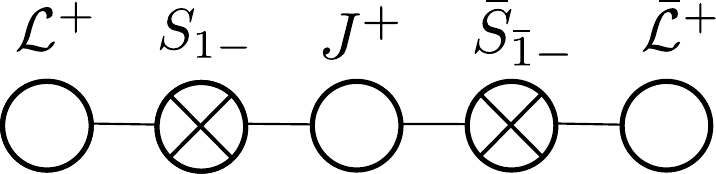}}}
	\caption{
			The ``Beauty'' Dynkin diagram of the 4D $\mathcal{N}=2$
			superconformal algebra, $\mathfrak{su}(2,2|2)$.
	}	\label{fig:beauty4d}
\end{figure}

As for any Lie superalgebra, there are several possible root systems to choose,
which we have found through a brute-force sweep in terms of the oscillator
representation of the generators, also collected in Appendix
\ref{app:oscillator-construction}.  The preferred Dynkin diagram for our
analysis is again associated with the maximal compact subgroup in equation
\eqref{4d-beauty-generators}, the ``Beauty'', and it is depicted in Figure
\ref{fig:beauty4d}. In terms of the generators, we have the triplets
\begin{gather}\label{4d-Beauty-roots}
		E_I^+ = (\mathcal{L}^+, S_{1-}, J^+, \bar{S}_{\bar{1}-}, \bar{\mathcal{L}}^+)\,,\qquad
		E_I^- = (E_I^+)^\dagger\,,\nonumber\\
		H_I = (\mathcal{L}^{(3)}, \frac{1}{2}(\mathcal{D} - J^{(3)}+\mathcal{L}^{(3)}+\frac{1}{2}U), J^{(3)}, \frac{1}{2}(\mathcal{D}-J^{(3)}+\bar{\mathcal{L}}^{(3)}-\frac{1}{2}U), \bar{\mathcal{L}}^{(3)})_I \,.
\end{gather}
We have written $H_2, H_4$ with respect to the generators giving rise to the
usual quantum numbers:
\begin{gather}
		\mathcal{L}^{(3)}\left|\Omega\right> = \ell\left|\Omega\right>\,,\qquad
		\bar{\mathcal{L}}^{(3)}\left|\Omega\right> = \bar{\ell}\left|\Omega\right>\,,\qquad
		\mathcal{D}\left|\Omega\right> = \Delta\left|\Omega\right>\,,\\
		J^{(3)}\left|\Omega\right> = j\left|\Omega\right>\,,\qquad
		U\left|\Omega\right> = j\left|\Omega\right>\,,
\end{gather}
with $U$ being associated with the $\mathfrak{u}(1)_r$ Abelian factor of the
R-symmetry. These generators satisfy the Serre--Chevalley relations
in equation \eqref{Serre--Chevalley}, with the Cartan matrix
\begin{equation}\label{4d-beauty-cartan}
		\mathcal{M}_{IJ} =
		\begin{pmatrix}
			2 & +1 & 0 & 0 & 0\\
			-1 & 0 & -1 & 0 & 0\\
			0 & -1 & 2 & -1 & 0\\
			0 & 0 & -1 & 0 & -1\\
			0 & 0 & 0 & +1 & 2
	\end{pmatrix}\,,
\end{equation}
The symmetrized Cartan matrix is given by $\mathcal{A}_{IJ} =
(\mathcal{M} \Lambda)_{IJ}$ with $\Lambda=\text{diag}(1,-1,-1,-1,1)$.

The conformal matter operators descend from the 6D parent theory: they are
half-BPS and uncharged under the Abelian factor of the R-symmetry, $r=0$. A
notable difference from the 6D case is that the shortening condition now leads to a
conformal dimension set by $\Delta^{4D}_{X}=j$ rather than the one given in
equation \eqref{Delta_CM}. The weight vector with respect to this choice of roots for any conformal matter
follows from the definition of the Cartan generators, see equation
\eqref{4d-Beauty-roots}:
\begin{equation}\label{weight-CM-4d}
		w = (0,0,j,0,0)\,,
\end{equation}
where the R-charge is given in Table \ref{tab:conformal-matter-operator}.

As already mentioned, the structure of integrability extends from the R-symmetry
sector to the full super-spin chain. Thus, we can also use the associated Bethe ansatz
equations for this super-spin chain, as in equation
\eqref{osp-Bethe-Ansatz-open}, where we now use the 4D symmetrized Cartan
matrix and the weight vector given in equation \eqref{weight-CM-4d}.

\section{Anomalous Dimensions and the Bethe Ansatz}\label{sec:BETHE}

Having established that we have an integrable super-spin chain which governs
operator mixing in a large R-charge sector of our 6D and 4D SCFTs, we now
extract the corresponding operator scaling dimensions of this system.  In
principle, all of this information is captured by solutions to the
corresponding Bethe ansatz equations, but in practice it can prove difficult to
explicitly solve for the rapidities and energies of this system.

A first cross-check is that our proposed answers are compatible with general
unitarity constraints on the representation theory of superconformal field
theories. Along these lines, recall that for 6D SCFTs, we imposed on physical
grounds an overall constraint on the net momentum of spin chain excitations
\cite{Baume:2020ure}:
\begin{equation}\label{decoupling-condition-6d}
		e^{i \sum_j (p^{I=5}_\perp)_j} = \prod_j \frac{u_j^{I=5} + \frac{i}{2}w^{I=5}}{u_j^{I=5} - \frac{i}{2}w^{I=5}} = 1.
\end{equation}
This constraint is required from a ``bottom up'' perspective as well. To see
why, suppose to the contrary that we had \textit{not} imposed such a
constraint. In this situation, we would simply admit all solutions to the Bethe
ansatz.  However, not all such solutions to the Bethe ansatz would lead to
physical solutions in the 6D SCFT.  Indeed, when a state acquires an anomalous
dimension, several short, protected, multiplets present in the free field limit
recombine into a larger long superconformal multiplet, denoted
$L[\ell_1,\ell_2,\ell_3]^j_\Delta$, where the quantum numbers refer
to those of the superconformal primary. In unitary theories, the conformal
dimension of long multiplets is bounded from below, and in six dimensions must
satisfy \cite{Cordova:2016xhm}:
\begin{equation}\label{long_multiplet_bound}
    L[\ell_1,\ell_2,\ell_3]^j_\Delta:\qquad\Delta > 2j + \frac{1}{2}(\ell_1 + 2\ell_2 + 3\ell_3)+ 6\,.
\end{equation}
It is easy to convince oneself that the shift from six leads to spurious
operators in the limit we are considering. For instance, for a single
excitation of the R-charge operator, $J^-$, the total R-charge is decreased by
two, therefore $j=L-2$, but the free-field dimension of that operator remains
$\Delta_0=2L$, and we have $\Delta = 2L + \gamma < 2(L-2) + 6$ which, since
$\gamma$ is small by assumption, violates unitarity. Imposing the net momentum
constraint in equation \eqref{decoupling-condition-6d}, we observe that the
unitarity bound is always respected.

The number of impurities is also constrained by superconformal symmetry. Due to
the structure of our choice of basis, some of the lowering operators $E_I^-$
annihilate the vacuum, $\vert 0 \rangle$ at a given site, and there is therefore
a hierarchy between the excitation numbers $n^I$ for each simple root. For instance,
due to the half-BPS nature of $X^L$, $E_4^-=Q^{1+}$ can only be used if $J^-$
has been used at least once, and we must have $n^4\leq n^5$. A similar analysis
for the other roots shows that
\begin{equation}\label{excitation-number-condition}
		0\leq n^3\leq n^2\leq n^1\leq n^4\leq n^5\,.
\end{equation}

Despite the two constraints in equations \eqref{decoupling-condition-6d} and
\eqref{excitation-number-condition}, finding solutions to the Bethe ansatz for
an arbitrary number of impurities is no simple task. While recent efforts to
use more modern computational algebro-geometric techniques have had some
success when applied to $\mathcal{N}=4$ super Yang--Mills \cite{Jiang:2017phk},
solving the Bethe ansatz is a notoriously difficult problem (see, however,
\cite{Marboe:2016yyn}).

While the general problem of extracting operator scaling dimensions appears
quite challenging, there are some closed subsectors where it is possible to
make further analytic progress. This includes the $\mathfrak{su}(2)_{R}$ closed
subsector studied in \cite{Baume:2020ure} and reviewed in section
\ref{sec:REVIEW}, as well as the $\mathfrak{sl}(2)$ subsector obtained
from lightcone derivatives of conformal matter. Additionally, ``fermionic''
excitations as captured by a $\mathfrak{su}(1|1)$ subalgebra form another
closed subsector. In each of these cases, there are only a limited number of
possible excitations, and the full (symmetrized) Cartan matrix of
$\mathfrak{osp}(6,2|1)$ appearing in the Bethe ansatz equations collapses to a
single parameter.  Even in this case, extracting closed form analytic
expressions is in general difficult, but in the special case of two impurities,
we can indeed extract analytic formulae. Our aim in the rest of this section
will be to illustrate the form of these closed subsectors in the two impurity
subsector, and to comment on the generalization to LSTs and 4D $\mathcal{N} =
2$ SCFTs, as appropriate.

Note that while the non-compact $\mathfrak{sl}(2)$ sector is highly excited
from the perspective of the Beauty, $P^{12}\left|0\right>\sim
Q^{1+}\mathcal{L}^1_2Q^{1+}J^-\left|0\right>$, it is however possible to use a super-Weyl
transformation to go to a root system for which $P^{12}$ is one of the ladder
operators, $E^-_I$. This transformation leads to the distinguished Dynkin
diagram, depicted in Figure \ref{fig:distinguished_6D}. A disadvantage of this
choice is that beyond a few impurities, the decoupling condition in equation
\eqref{decoupling-condition-6d} is obscured.

\begin{table}
	\centering
	\begin{tabular}{l|c|c|c|c|c|c}
	subsector 	     & $E^+$      & $E^-$      & $H$              & Impurity         & $w$ & $\mathcal{A}$\\\hline
	$\mathfrak{su}(2)_R$   & $J^+$    & $J^-$    & $J^{(3)}$        & $Y^\dagger$      & $+1$                   & $(2)$ \\
	$\mathfrak{su}(1|1)$ & $S_{1+}$ & $Q^{1-}$ & $\mathcal{L}^1_1 + J^{(3)}$   & $\psi^1$         & $+2$                   & $(0)$\\
	$\mathfrak{sl}(2)$   & $K_{12}$ & $P^{12}$ & $-(\mathcal{L}^1_1 + \mathcal{L}^2_2)$ & $DX$ & $-2$                   & $(2)$
	\end{tabular}
	\caption{Closed rank-one subsectors generated by triplets $(E^+,E^-,H)$ and
			the associated type of impurities in six dimensions, with their Cartan matrix and weight vector of the hypermultiplet.
	}
	\label{tab:rank-one-subsectors}
\end{table}

\subsection{Rank One Subsectors}

When considering any of the rank-one closed subsectors, the Bethe ansatz with
$n$ impurities reduces to a variation of that of the Heisenberg XXX$_{s = \frac{w}{2}}$ spin chain:
\begin{equation}\label{eqn:bethe-rank-one}
	\left(  \frac{u_{j}+\frac{i}{2}w}{u_{j}-\frac{i}{2}w}\right)^{2L}
	=
	\prod_{k\neq j}^{n}
	\frac{u_{j} - u_{k} + \frac{i}{2}\mathcal{A}}{u_{j} - u_{k} - \frac{i}{2}\mathcal{A}}\,
	\frac{u_{j} + u_{k} + \frac{i}{2}\mathcal{A}}{u_{j} + u_{k} - \frac{i}{2}\mathcal{A}}\,,
\end{equation}
where $w, \mathcal{A}$ are now integers, and summarized in Table
\ref{tab:rank-one-subsectors} for each subsector.

\paragraph{The Fermionic $\mathfrak{su}(1|1)$ Sector:} The simplest case is that of fermionic
excitations, as the RHS collapses to one. We can then treat all conformal
matter cases uniformly:
\begin{equation}\label{Bethe-fermionic}
		\left(\frac{u_{j}+\frac{i}{2}w}{u_{j}-\frac{i}{2}w}\right)^{2L} = e^{2ip_j L}= 1\,.
\end{equation}
The Bethe ansatz therefore implies that all the fermionic impurities are
uncorrelated and the excitations correspond to plane waves traveling
freely along the spin chain with momenta
\begin{equation}\label{sol-su(1|1)}
		p_j = \frac{\pi m }{L}\,,\qquad
		m = 0, 1,\cdots,L-1
\end{equation}
In the two-impurity sector, the decoupling condition imposes $p=p_1= -p_2$, and
we find that the anomalous dimensions are given by
\begin{equation}
	\gamma = (\Delta-\Delta_0) = \lambda_G \frac{8}{\vert w \vert}\sin^2(\frac{p}{2})\,,
\end{equation}
with $w= 1+j$ for any of the R-charges given in Table
\ref{tab:conformal-matter-operator}.

Similar considerations apply for our 4D $\mathcal{N} = 2$ SCFTs. There are two
equivalent $\mathfrak{su}(1|1)$ sectors associated with the chiral/anti-chiral
fermions of the $\mathcal{N}=2$ hypermultiplets generated by $S_{1-}$ or
$\bar{S}_{\bar{1}-}$. For any of the 4D conformal matter quivers, we therefore
obtain the same functional form for the Bethe ansatz equations, substituting
for the proper weights $w=\frac{1}{2}(1+j)$, for any of the R-charges given in
Table \ref{tab:conformal-matter-operator}. As an additional comment, note that
for 4D $\mathcal{N} = 2$ SCFTs, there is no overall momentum constraint on the
excitations as in equation \eqref{decoupling-condition-6d}. Since, however, the
equations in this sector completely decouple, this poses no obstacle to
obtaining similar expressions for the rapidities.

\paragraph{The $\mathfrak{sl}(2)$ and $\mathfrak{su}(2)_R$ Subsectors:} It
turns out that these two cases can be analyzed together. Indeed, the Cartan
matrix is the same, and we have the case of an XXX$_{s=\frac{w}{2}}$ Heisenberg
spin chain, where we now allow for $s$ to be a negative half-integer in the
case of the non-compact $\mathfrak{sl}(2,\mathbb{R})$ algebra. In the
two-impurity sector $u=u_1=-u_2$, and the Bethe ansatz simplifies to
\begin{equation}\label{bethe-sl2-su2}
		\left( \frac{u+\frac{i}{2}w}{u-\frac{i}{2}w} \right)^{2L} = - \frac{u + \frac{i}{2}}{u - \frac{i}{2}}\,.
\end{equation}
For the case $w=1$, we find again the case of the $\mathfrak{su}(2)_R$ spin
chain associated with the hypermultiplet, namely A-type conformal matter.
In that case, the momenta are found to be
\begin{equation}
		p = p_1 = -p_2=\frac{\pi(2m+1)}{2L-1}\,,\qquad m=0,\cdots, L-1\,,
\end{equation}
and the anomalous dimension is given by
\begin{equation}
		\gamma = (\Delta-\Delta_0) = 8 \lambda_A  \sin^2 \left(\frac{\pi(2m+1)}{4L-2} \right)\,,
\end{equation}
with $\lambda_A$ for a 6D SCFT with A-type conformal matter as given in Section
\ref{sec:REVIEW}.  For D- and E-type conformal matter, there is no such
simplification. We can, however, estimate the anomalous dimension by performing
a formal large-$L$ and large-$w$ expansion and solving the Bethe ansatz in equation
\eqref{bethe-sl2-su2} in perturbation theory (see reference
\cite{Baume:2020ure}):
\begin{equation}
		p = \frac{\pi(2m+1)}{2L}\left(1 + \frac{1}{2wL} + \cdots \right)\,,\qquad m=0,\cdots, L-1\,.
\end{equation}
The anomalous dimension is then
\begin{equation}
		\gamma = (\Delta-\Delta_0) = \frac{8 \lambda_{G}}{\vert w \vert} \sin^2(p)\,,
\end{equation}
with $\lambda_{G}$ specified as in Section \ref{sec:REVIEW}.
The weight vector is given by $w=j$ for the $\mathfrak{su}(2)_R$ sector, and
$w=-2j$ for the $\mathfrak{sl}(2)$ sector associated with lightcone
derivatives.

The same reasoning also applies to the 4D light-cone derivative sector generated
by $P^{1\bar{1}}$ or the R-symmetry sector. The only difference is that the
weight vector is now given by $w=-j$ for the former, and $w=j$ for the latter.
In that sense, the formal large-$w$ expansion converges faster for the 6D spin
chain. Similar considerations clearly hold for the LST sector, since we only
need to work with periodic boundary conditions instead. For a related
discussion in the case of 4D SCFTs, see reference \cite{Baume:2020ure}.

\section{Conclusions}\label{sec:CONC}

Higher-dimensional conformal field theories are important in the study of
strongly coupled quantum field theories.  In spite of their importance, it has
proven remarkably difficult to extract microscopic details on the spectrum of
local operators in such theories. Leveraging their partial tensor branch
description in terms of generalized quiver gauge theories, in this paper we
have shown that there are large R-charge operators governed by an
$\mathfrak{osp}(6,2|1)$ super-spin chain.  On the tensor branch the ground
state of the spin chain is specified as a gauge invariant product of
particular bifundamental conformal matter operators. The superconformal
descendant of a bifundamental conformal matter operator corresponds to
an excitation of the spin chain. This gives rise to a whole class of operators
which close under operator mixing (as specified by the dilatation operator). To
leading order in an inverse R-charge expansion, the dilatation operator
specifies a Hamiltonian with nearest neighbor interactions. Integrability of
the $\mathfrak{su}(2)_R$ R-symmetry subsector naturally extends to derivative
insertions governed by an $\mathfrak{sl}(2)$ subsector, and this in turn lifts
to an integrable structure on the full $\mathfrak{osp}(6,2|1)$ super-spin chin.
The corresponding Bethe ansatz equations then implicitly determine the operator
scaling dimensions for this system, which can be analytically solved in certain
closed subsectors with a low number of impurities. This methodology applies to
ADE conformal matter, as well as related 6D LSTs and 4D $\mathcal{N} = 2$
SCFTs. In the remainder of this section, we discuss some avenues for further
investigation.

In the case of D- and E-type conformal matter, there is a natural
interpretation in terms of ``bound states'' built up from A-type conformal
matter states. Using our analysis of the $\mathfrak{osp}(6,2|1)$ super-spin
chain, it should be possible to determine the corresponding binding energies /
scaling dimensions associated with A-type conformal matter excitations and their
associated D- and E-type counterparts.

Although our primary emphasis has been on the operator content of the
associated spin chain, there is clearly a great deal of geometric structure, as
captured by the corresponding F-theory realization of the theory. Indeed, in
reference \cite{Heckman:2014qba}, the F-theory geometry for conformal matter
was directly used to extract the scaling dimensions for half-BPS states. Given
this, it is tempting to speculate that the spectral curve of the algebraic
Bethe ansatz is directly visible in the F-theory geometry. It would be
interesting to investigate this possibility further.

A general question is whether the integrable structure observed here
persists beyond leading order in perturbation theory.
The answer for related 4D $\mathcal{N} = 2$ theories was that such integrable
structures can break down at two-loop order \cite{Gadde:2010zi} (see also
\cite{Nunez:2018ags, Filippas:2019puw}). The operator sector considered in this
paper is somewhat different, and the very fact that it is so close to the
orbifold of $\mathcal{N} = 4$ super Yang--Mills
\cite{Beisert:2005he} suggests that integrability may be more robust in these
systems. Investigating whether integrability persists once longer range
interaction terms are taken into account would no doubt be worthy of further study.

While our analysis has focused on a spin-chain characterization of
operator mixing in 6D SCFTs, we have also made use of the holographic dual to
provide additional guidance in the study of this system. It would of course be
interesting to directly construct the operators in the gravity dual. These are generated by
the fluctuations of M2-branes suspended along the
great arc of $S^4 / \Gamma_{ADE}$ which end at the north and south pole orbifold
fixed points. This would also provide a cross-check on leading order integrability
in an inverse R-charge expansion.

Finally, compactification of a 6D SCFT on a $T^2$ provides a general template
for realizing 4D theories, as well as dualities amongst such QFTs. From this
perspective, it is natural to ask what becomes of our spin-chain sector when we apply
a duality transformation in the lower-dimensional theory. This would appear to take us to a
seemingly completely different class of 4D operators. Analyzing this map holds out the prospect of
obtaining microscopic details on the nature of duality.

\section*{Acknowledgements}

We thank S.~Ekhammar, P.~Koreteev, and D.~Volin for helpful discussions, as well as F.~Hassler and J.A.~Minahan for helpful correspondence.
FB is supported by the Swiss National Science Foundation (SNSF), grant number P400P2\_194341.
The work of JJH is supported by the DOE (HEP) Award DE-SC0013528. CL acknowledges support from DESY (Hamburg, Germany),
a member of the Helmholtz Association HGF.

\newpage

\appendix
\section{Oscillator Construction}\label{app:oscillator-construction}

In this appendix, we present a convenient construction of the 4D and 6D
superconformal algebras in terms of a collection of harmonic oscillators.
Aside from its utility in constructing various representations, it also
provides a physical interpretation of operators as
quasi-particle excitations of the spin chain. Since we
are interested in the application to spin chains, we take one copy of the
oscillators for each spin chain site. We begin with the oscillator construction
for 6D $\mathcal{N}=(1,0)$ SCFTs and then turn to 4D $\mathcal{N} = 2$ SCFTs.

In what follows, we shall find it convenient to make use of the Jordan
decomposition of a \emph{non-compact} Lie (super)algebra:
\begin{equation}
	\mathfrak{g=g}^{-}\oplus\mathfrak{g}_{0}\oplus\mathfrak{g}^{+}\,.
\end{equation}
Here, $\mathfrak{g}_{0}$ is the maximal \emph{compact} subalgebra,
$\mathfrak{g}^{-}$ and $\mathfrak{g}^{+}$ are exclusively generated by lowering
and raising operators, and satisfy the three-grading
\begin{equation}\label{three-grading}
	[\mathfrak{g}_0, \mathfrak{g}^\pm]\subset\mathfrak{g}^\pm\,,\qquad
	[\mathfrak{g}^+,\mathfrak{g}^-]\subset\mathfrak{g}_0\,.
\end{equation}
We note that while similar to the Jordan decomposition of a non-compact
algebra, this is not the same as the usual Cartan decomposition into eigenspaces
with respect to the adjoint action of the Cartan generators. Here, some of the lowering
and raising generators are contained in $\mathfrak{g}_{0}$ in addition to the
Cartan subalgebra.\footnote{Our convention is such that $\mathfrak{g}^{+}$ is
		associated with \emph{annihilation} bi-oscillators, which is different
		from e.g., that of reference \cite{Gunaydin:1981yq}. Our choice is made
so that elements of $\mathfrak{g}^{+}$ are associated with \emph{positive}
roots, and therefore raising operators with respect to the root system rather
than the oscillator vacuum.} An advantage of this decomposition is that, while
the irreducible representations of the (super-)conformal algebras are
generically infinite-dimensional, those of their maximal compact subgroup,
$\mathfrak{g}_0$, are conversely always finite-dimensional. Moreover, if an
irreducible representation of $\mathfrak{g}_0$ is such that it is annihilated
by all elements of $\mathfrak{g}^+$, it induces an irreducible representation
of $\mathfrak{g}$, where the infinite tower of descendants is reached by
successive action of elements of $\mathfrak{g}^-$ \cite{Gunaydin:1981yq,
Bars:1982ep}.

\subsection{6D \texorpdfstring{$\mathcal{N}=(1,0)$}{N=(1,0)} SCFTs: \texorpdfstring{$\mathfrak{osp}(6,2|1)$}{osp(6,2|1)}}\label{app:6Doscillators}

The 6D superconformal algebra is $\mathfrak{osp}(6,2|1)$, which has the
following Jordan decomposition \cite{Gunaydin:1990ag, Gunaydin:1999ci,
Gunaydin:1981yq}:
\begin{equation}\label{Jordan-decomposition_6d}
	\mathfrak{osp}(6,2|1) = \mathfrak{g}^{-}
	\oplus\mathfrak{g}_{0}\oplus\mathfrak{g}^{+}\,,\qquad\mathfrak{g}_{0} =
	\mathfrak{u}(4|1)\,,
\end{equation}
At the non-supersymmetric level, there is a similar decomposition for the
ordinary conformal algebra, $\mathfrak{so}(6,2)$, with respect to its own
maximal compact subalgebra, $\mathfrak{u}(4)$. The
$\mathfrak{su}(4)=\mathfrak{so}(6)$ part of this subalgebra corresponds to the
Lorentz group, while the Abelian factor is related to dilatations. We stress
that the decomposition in equation \eqref{Jordan-decomposition_6d} is in contradistinction
with the usual decomposition in terms of the bosonic subalgebra
$\mathfrak{osp}(6,2|\mathcal{N})\to\mathfrak{so}(6,2)\oplus \mathfrak{sp}(\mathcal{N})_R$.
Moreover, as pointed out in \cite{Gunaydin:1999ci} the choice of spacetime
signature is not particularly significant.

We can then define two pairs of bosonic oscillators, transforming in the
(anti-)fundamental representation of $\mathfrak{su}(4)$ (or equivalently the spinor
representation of $\mathfrak{so}(6)$): $a_{i}$ and $b_{i}$, where $i =
1,\cdots, 4$, as well as a pair of fermionic oscillators $(\alpha, \beta)$
charged under the R-symmetry Cartan subalgebra of $\mathfrak{sp}(1)_{R}=\mathfrak{su}(2)_{R}$. They
can be combined into a pair of superoscillators transforming in irreducible
representations of $\mathfrak{u}(4|1)$,
\begin{gather}
\xi_{A} = (a_{i}, \alpha)\,,\qquad\eta_{A} = (b_{i}, \beta)\,,\\
\xi^{A} = \xi_{A}^{\dagger}= (a^{i}, \alpha^{\dagger})\,,\qquad\eta^{A} =
\eta_{A}^{\dagger}= (b^{i}, \beta^{\dagger})\,,
\end{gather}
satisfying canonical (anti-)commutation relations:
\begin{equation}\label{oscillators-commutation-relations}
		[\xi_{A}(l),\xi^{B}(l')\} = \delta^{B}_{A}\delta_{l,l'} = [\eta_{A}(l),\eta^{B}(l')\}\,,
\end{equation}
and zero in all other cases, with $[\cdot,\cdot\}$ denoting the
anti-commutators for two fermionic oscillators and commutators otherwise. For
ease of notation, oscillators with raised indices denote creation operators,
e.g. $a^i=(a_i)^\dagger$. We have also allowed for multiple copies
$l,l'=1,\cdots,L$ associated with each site of the spin chain. Each of the
superconformal generators can then be expressed as bi-oscillators satisfying
the three-grading in equation \eqref{three-grading} \cite{Gunaydin:1990ag,Gunaydin:1999ci}:
\begin{align}\label{eqn:gens-6D}
		\mathfrak{g}^{+}: \qquad A_{AB}  &  = \xi_{A}\eta_{B}-\eta_{A} \xi_{B} = K_{ij}\oplus J^{+}\oplus S_{i+}\,;\nonumber\\
		\mathfrak{g}^{-}: \qquad A^{AB}  &  = A_{AB}^{\dagger}= \eta^{B}\xi^{A}-\xi^{B}\eta^{A} = P^{ij}\oplus J^{-}\oplus Q^{i-}\,;\\
		\mathfrak{g}_{0}: \qquad M^{A}_{B}  &  = \xi^{A}\xi_{B} + (-1)^{(\text{deg }A)(\text{deg } B)}\eta_{B}\eta^{A} = \mathcal{D}\oplus\mathcal{L}^{i}_{j} \oplus J^{(3)} \oplus
\widetilde{Q}^{i+}\oplus\widetilde{S}_{i-}\,.\nonumber
\end{align}

The Lorentz generators, $\mathcal{L}^i_j$, and the dilatation operator,
$\mathcal{D}$, are by construction part of the maximal compact algebra,
$\mathfrak{g}_{0}$. With the Cartan generator of the R-symmetry, $J^{(3)}$,
they generate the bosonic part of $\mathfrak{g}_0$, and are defined as:
\begin{gather}
		\mathcal{L}^{i}_{j} = a^{i}\cdot a_{j} + b_{j}\cdot b^{i}\,,\qquad
		\mathcal{D} = \frac{1}{2}\sum_{i=1}^{4}M^{i}_{i} = 2L + \frac{1}{2}(N_{a} + N_{b})\,,\\
		J^{(3)} = -(\alpha^{\dagger}\cdot\alpha- \beta\cdot\beta^{\dagger}) = L - (N_{\alpha}+ N_{\beta})\,.
\end{gather}
where $a^{i}\cdot a_{j} =\sum_{l=1}^{L} a^{i}(l)a_{j}(l)$, and $N_{a} =
\sum_{i=1}^{4} a^{i}\cdot a_{i}$ is the number operator for oscillators of type
$a$, defined similarly for $b,\alpha,\beta$.

The special conformal transformations $K_{ij}$, translations $P^{ij}$,
as well as the R-charge ladder operators, $J^{\pm}$, are part of
$\mathfrak{g}^{\pm}$:\footnote{The oscillator representation uses a bispinor
		notation, e.g., $K_{ij} \sim\Gamma^{\mu}_{ij}K_{\mu}\,,\, P^{ij}\sim
\widetilde{\Gamma}^{ij}_\mu P^\mu$.}
\begin{align}
\mathfrak{g}^{+}:  &  \qquad K_{ij}= a_{i}\cdot b_{j} - a_{j}\cdot
b_{i}\,,\qquad J^{+} = \alpha\cdot\beta\,;\\
\mathfrak{g}^{-}:  &  \qquad P^{ij}= a^{i}\cdot b^{j} - a^{j}\cdot
b^{i}\,,\qquad J^{-} = \beta^{\dagger}\cdot\alpha^{\dagger}\,.
\end{align}

Finally, the fermionic generators correspond to the 8+8 super(conformal)
charges:
\begin{align}
Q^{i+}  &  =a^{i}\cdot\alpha+b^{i}\cdot\beta\,, & Q^{i-}  &  =a^{i}\cdot
\beta^{\dagger}-b^{i}\cdot\alpha^{\dagger}\,,\\
S_{i-}  &  =a_{i}\cdot\alpha^{\dagger}+b_{i}\cdot\beta^{\dagger}\,, & S_{i+}
&  =a_{i}\cdot\beta-b_{i}\cdot\alpha\,,
\end{align}
The sign is chosen to correspond to the R-charge, e.g. $[J^{(3)},Q^{i\pm}]=\pm
Q^{i\pm}$, and as expected, they transform as doublets
of $\mathfrak{su}(2)_{R}$.

Finally, the $\mathfrak{osp}(6,2|1)$ quadratic Casimir is given by
\begin{equation}\label{quadratic-casimir-osp}
		\begin{aligned}
			C_{\mathfrak{osp}(6,2|1)}=  &  H_{I}\mathcal{A}^{IJ}H_{J}+ \sum_{i<j} \left(\{\mathcal{L}^i_j,\mathcal{L}^j_i\} + \{K_{ij}, -P^{ij}\}\right)\\
			&   -\sum_i\left([S_{i+},Q^{i-}] + [S_{i-},Q^{i+}]\right) - 2 \{J^+,J^-\}\,,
		\end{aligned}
\end{equation}
where the Cartan generators, $H_I$, and the inverse of the symmetrized Cartan
matrix, $\mathcal{A}^{IJ}=(\mathcal{A}^{-1})_{IJ}$, are defined with respect to
the Beauty root system, see Section \ref{sec:beauty-6D}.

\paragraph{Commutation relations:} for the reader's convenience, we give here
the list of all commutation relations amongst the generators of
$\mathfrak{osp}(6,2|1)$:

\begin{itemize}
	\item
		Relations involving Lorentz generators $\mathcal{L}^i_j$:
		\begin{gather}
				[\mathcal{L}^i_j, \mathcal{L}^k_l] = \delta^k_j \mathcal{L}^i_l - \delta^i_l \mathcal{L}^k_j\,,\nonumber\\
			[\mathcal{L}^i_j, K_{kl}] = \delta^i_lK_{jk} - \delta^i_k K_{jl}\,,\quad
			[\mathcal{L}^i_j, P^{kl}] = \delta^k_jP^{il} - \delta^l_j P^{ik}\,,\\
			[\mathcal{L}^i_j, J] = 0\,,\qquad
			[\mathcal{L}^i_j, Q^{k\pm}] = \delta^k_j Q^{i\pm}\,,\qquad
			[\mathcal{L}^i_j, S_{k\pm}] = -\delta^i_k S_{j\pm}\,.\nonumber
		\end{gather}
	\item
		Relations involving special conformal transformations and translations $K_{ij}$ and $P^{ij}$:
		\begin{gather}
			[K_{ij}, P^{kl}] = \delta_i^k \mathcal{L}_j^l + \delta_j^l \mathcal{L}^k_i - \delta_i^l \mathcal{L}_j^k - \delta_j^k \mathcal{L}^l_i\,,\nonumber\\
			[K_{ij},Q^{k\pm}] = \mp(\delta_i^k S_{j\pm} - \delta^k_j S_{i\pm})\,,\qquad
			[P^{ij}, S_{k\pm}] = \mp(\delta^i_k Q^{j\pm} - \delta^j_k Q^{i\pm})\,,\qquad\nonumber\\
			[K,S] = 0 = [P,Q]\,,\qquad [K,J] = 0 = [P,J]\,.\nonumber
		\end{gather}
	\item Relations involving R-symmetry generators $J^\pm, J^{(3)}$:
	\begin{gather}
		[J^+, J^-] = J^{(3)}\,,\qquad [J^{(3), J^\pm}] = \pm2 J^{\pm}\,,\nonumber\\
		[J,\mathcal{L}] = 0\,,\qquad [J,P] = 0 = [J,K]\,,\nonumber\\
		[J^{(3)}, Q^{i\pm}] = \pm Q^{i\pm}\,,\qquad
		[J^{(3)}, S_{i\pm}] = \pm S_{i\pm}\,,\nonumber\\
		[J^\pm,Q^{i\pm}] = 0 = [J^\pm,S_{i\pm}]\,,\\
		[J^{\pm}, Q^{i\mp}] = + Q^{i\pm}\,,\qquad
		[J^{\pm}, S_{i\mp}] = - S_{i\pm}\,.\qquad
	\end{gather}
	\item Anti-commutation relations for the fermionic generators:
	\begin{gather}
		\{Q^{i+}, S_{j-} \} = \mathcal{L}^i_j -  \delta^{i}_{j} J^{(3)}\,,\qquad
		\{Q^{i-}, S_{j+}  \} = \mathcal{L}^i_j + \delta^i_j J^{(3)}\,,\nonumber\\
		\{Q^{i+},S_{j+} \} = - 2\delta^i_j J^+\,,\qquad
		\{Q^{i-},S_{j-}\} = -2\delta^i_j J^-\,,\\
		\{Q^{i+},Q^{j-}\} = - P^{ij}\,,\qquad
		\{S_{i+}, S_{j-}\} = K_{ij}\,.\nonumber
	\end{gather}
	\item Commutation relations involving the dilatation operator:
	\begin{gather}
			[\mathcal{D}, K] = - K\,,\qquad
			[\mathcal{D}, P] = + P\,,\qquad
			[\mathcal{D}, J] = 0 = [\mathcal{D},\mathcal{L}]\,,\\
			[D, Q] = +\frac{1}{2}Q\,,\qquad
			[D, S] = -\frac{1}{2}S\,.
	\end{gather}
\end{itemize}

\paragraph{Representations of Superconformal Multiplets:} Conformal operators
are constructed out of the oscillator vacuum $\left|0\right>$, and are of the
form $\prod_{a,l} \xi^{A_a}(l)\eta^{B_a}(l)\left|0\right>$, where we allow for
both multiple sites and multiple excitations of the vacuum. It is easy to see
that the representations $\mathfrak{g}_0=\mathfrak{u}(4|1)$ are finite
dimensional, and if a highest weight state $\left|\Omega\right>$ of
$\mathfrak{g}_0$ is further annihilated by all elements of $\mathfrak{g}^+$, it
is moreover a highest weight state of $\mathfrak{osp}(6,2|1)$. This means that
any state $\left|\Omega\right>$ with these properties generates a full,
infinite-dimensional, representation of superconformal multiplets by successive
applications of elements of $\mathfrak{g}^-$.

The example of the 6D $\mathcal{N}=(1,0)$ free hypermultiplet is detailed in
Section \ref{sec:DYNKIN}. In principle, we can construct any
superconformal multiplet via the oscillator construction. For instance, there
is another free field multiplet in 6D $\mathcal{N}=(1,0)$: the tensor multiplet,
$C[0,0,0]^{j=0}_{\Delta=2}$. Its primary is a scalar uncharged under
the R-symmetry, and it contains the self-dual tensor characteristic of 6D SCFTs
as superconformal descendants. In terms of oscillators, the scalar primary is
represented by $(\alpha^\dagger+\beta^\dagger)\left|0\right>$. It is
straightforward to check that it has the correct quantum numbers, and its
descendants indeed form a $C[0,0,0]^0_2$ superconformal multiplet.
However, when turning on interactions, the free tensors will recombine into
strongly-coupled long multiplets, whose microscopic dynamics are beyond the
scope of this work.

\begin{figure}[t!]
\centering
\begin{subfigure}[t]{.49\linewidth}
		\centering
		\begin{minipage}{.1cm}
			\vfill
		\end{minipage}
		{\vspace{0.4cm}\resizebox{0.49\textwidth}{!}{\includegraphics{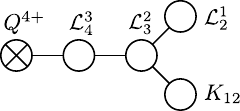}}}
		\caption{Distinguished diagram}
		\label{fig:distinguished_6D}
	\end{subfigure}
\hfill\begin{subfigure}[t]{.49\linewidth}
		\centering
		{\vspace{-1.3cm}\resizebox{0.49\textwidth}{!}{\includegraphics{figures/beauty_6D.pdf}}\vspace{.8cm}}
		\caption{The ``Beauty''}
		\label{fig:beauty_6D}
	\end{subfigure}
\hfill\begin{subfigure}[t]{.49\linewidth}
		\centering
		{\vspace{0.4cm}\resizebox{0.49\textwidth}{!}{\includegraphics{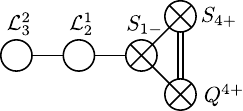}}}
		\caption{}
		\label{fig:3nodes}
	\end{subfigure}
\hfill\begin{subfigure}[t]{.49\linewidth}
		\centering
		{\vspace{0.4cm}\resizebox{0.49\textwidth}{!}{\includegraphics{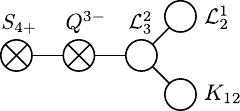}}}
		\caption{}
		\label{fig:distinguished_2nodes_1_6D}
	\end{subfigure}
\hfill\begin{subfigure}[t]{.49\linewidth}
		\centering
		{\vspace{0.4cm}\resizebox{0.49\textwidth}{!}{\includegraphics{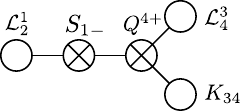}}}
		\caption{}
		\label{fig:distinguished_2nodes_2_6D}
	\end{subfigure}
\caption{
		\label{fig:dynkin-6d}
		Topologically distinct Dynkin diagrams of the 6D
		superconformal algebra, $\mathfrak{osp}(6,2|1)$, with each node labeled
		by the associated generator, $E^+_I$. Fermionic nodes are denoted by a
		cross, and the representative is chosen such that $X$ is the
		highest weight state of its representation.
}
\end{figure}

\paragraph{Root Systems and Dynkin Diagrams:} As explained in section
\ref{sec:DYNKIN}, the root system of a Lie superalgebra is not unique and leads
to multiple possible Dynkin diagrams. We have scanned over every possibility
through the following procedure. We first look for
$\text{rank}(\mathfrak{osp}(6,2|1))=5$ triplets $(E^+_I,E^-_I, H_I)$ satisfying
the Serre--Chevalley relations \eqref{Serre--Chevalley}. The associated Dynkin
diagram is then well defined if every root can be written as integer
combinations of the simple roots. This procedure is easily implemented on a
computer, and results in the five topologically-different Dynkin diagrams
collected in Figure \ref{fig:dynkin-6d}. There are many equivalent
representatives, and in each case it is chosen such that $X$ is the highest
weight state the induced representation. The distinguished Dynkin diagram,
extending the conformal group with a single fermionic node, is shown in Figure
\ref{fig:distinguished_6D}, while the ``Beauty'', which makes the maximal
compact subalgebra, $\mathfrak{u}(4|1)$, manifest and is the choice used in this work, is given in Figure
\ref{fig:beauty_6D}.

\subsection{4D \texorpdfstring{$\mathcal{N}=2$}{N=2} SCFTs: \texorpdfstring{$\mathfrak{su}(2,2|2)$}{su(2,2|2)}}

A similar construction can also be implemented for 4D $\mathcal{N}=2$ SCFTs,
whose superconformal algebra is $\mathfrak{su}(2,2|2)$. The Jordan
decomposition is given by \cite{Gunaydin:1984fk, Gunaydin:1998jc,
Gunaydin:1998sw}:
\begin{equation}\label{Jordan-decomposition_4d}
	\mathfrak{su}(2,2|2)=\mathfrak{g}^{-}\oplus\mathfrak{g}_{0}\oplus
	\mathfrak{g}^{+}\,,\qquad\mathfrak{g}_{0}=\mathfrak{su}(2|1)\oplus
	\mathfrak{su}(2|1)\oplus\mathfrak{u}(1)\,.
\end{equation}
At the non-supersymmetric level, the decomposition of the conformal algebra,
$\mathfrak{su}(4,2)$, is given by $\mathfrak{su}(2)\oplus\mathfrak{su}(2)\oplus
\mathfrak{u}(1)$. As in six dimensions, this is different from the usual
decomposition $\mathfrak{su}(2,2|2)\to\mathfrak{so}(4,2)
\oplus\mathfrak{su}(2)_{R} \oplus\mathfrak{u}(1)_{r}$\,.

To construct a representation of each of the generators, we define two pairs of
bosonic annihilation operators transforming in the doublet of each
$\mathfrak{su}(2)$: $a_{i}$ and $b_{\bar{k}}$, where $i,\bar{k}=1,2$. We also
define a pair of fermionic oscillators $(\alpha,\beta)$, which will be
associated with the R-symmetry.

These operators can then be combined into two super-oscillators, transforming
in the fundamental representation of each $\mathfrak{su}(2|1)$ factors
appearing in $\mathfrak{g}_{0}$.
\begin{gather}
	\xi_{A}=(a_{i},\alpha)\,,\qquad \eta_{A}=(b_{\bar{k}},\beta)\,,\\
	\xi^{A}=\xi_{A}^{\dagger} = (a^{i},\alpha^{\dagger})\,,\qquad
	\eta^{A}=\eta_{A}^{\dagger}=(b^{\bar{k}},\beta^{\dagger})\,,
\end{gather}
The two indices $i,\bar{k}$ can be viewed as 4D spinor indices for each
chirality of $\mathfrak{so}(4)=\mathfrak{su}(2)\oplus\mathfrak{su}(2)$, such
that a Lorentz vector, e.g. $P_{\mu}$ is understood in this notation as a
bispinor $P_{i\bar{k}}\sim\sigma_{i\bar{k}}^{\mu}P_{\mu}$. Each oscillator
comes in multiple copies, e.g. $\alpha(l)\,,l=1,\cdots,L$, associated with each
site of the spin chain, and satisfy the usual canonical (anti-)commutation relations.
\begin{equation}
		[\xi_{A}(l),\xi^{B}(l')\} = \delta^{B}_{A}\,\delta_{l,l'} = [\eta_{A}(l),\eta^{B}(l')\}\,.
\end{equation}
The
generators of the superconformal algebra can then be written as bi-oscillators
in such a way that the three-grading in equation \eqref{three-grading} is
satisfied:
\begin{align}\label{eqn:gens-4D}
		\mathfrak{g}^{+}: \qquad A_{AB}  &  = \xi_{A}\cdot\eta_{B} = K_{i\bar{k}}\oplus J^{+}\oplus S_{i+}\oplus\bar{S}_{\bar{k}+}\,;\nonumber\\
		\mathfrak{g}^{-}: \qquad A^{BA}  &  = A_{AB}^{\dagger}= \eta^{B}\cdot\xi^{A} = P^{i\bar{k}}\oplus J^{-}\oplus Q^{i-}\oplus\bar{Q}^{\bar{k}-}\,;\\
		\mathfrak{g}_{0}\,: \qquad\, M^{A}_{B}\,  &  = \xi^{A}\cdot\xi_{B} \oplus \eta_{B}\cdot\eta^{A} = \mathcal{L}^{i}_{j} \oplus\bar{\mathcal{L}}^{\bar{k} }_{\bar{l}}\oplus J^{(3)}\oplus U\oplus Q^{i+}\oplus S_{i-} \oplus\bar{Q}^{\bar{k}+}\oplus\bar{S}_{\bar{k}-}\,,\nonumber
\end{align}
where the scalar product indicates a sum over all generations, e.g. $\eta
_{A}\cdot\xi_{B} =\sum_{l=1}^{L}\eta_{A}(l)\xi_{B}(l)$. Note that contrary to
the 6D case, the two types of superoscillators are uncorrelated in
$\mathfrak{g}_0$ reflecting the unitary, rather than orthosymplectic, structure
of $\mathfrak{su}(2,2|2)$.

The bosonic generators of $\mathfrak{g}_0$ contain the Lorentz generators, $\mathcal{L}^i_i$,
$\bar{\mathcal{L}}^{\bar{k}}_{\bar{l}}$, the dilatation operator,
$\mathcal{D}$, as well as the Cartan generators of the R-symmetry
$\mathfrak{u}(2)=\mathfrak{su}(2)_R\oplus\mathfrak{u}(1)_r$. They are realized
as bosonic combinations of creation and annihilation operators:
\begin{gather}
	\mathcal{L}_{j}^{i}=a^{i}\cdot a_{j}-\frac{1}{2}\delta_{j}^{i}a^{m}\cdot
	a_{m}\,,\qquad\bar{\mathcal{L}}_{\bar{l}}^{\bar{k}}=b^{\bar{k}}\cdot
	b_{\bar{l}}-\frac{1}{2}\delta_{\bar{l}}^{\bar{k}}b^{\bar{m}}\cdot b_{\bar{m}}\,,\\
	\mathcal{D}=\frac{1}{2}(a^{i}\cdot a_{i}+b_{\bar{k}}\cdot b^{\bar{k}})=\frac{1}{2}(N_{a}+N_{b})+L\,,\\
	J^{(3)}=-(\alpha^{\dagger}\alpha-\beta\beta^{\dagger})=-N_{\alpha}-N_{\beta}+L\,,\\
	U=(N_{a}-N_{b})+2(N_{\alpha}-N_{\beta})\,,
\end{gather}
with $N_{a}=a^{i}\cdot a_{i}\,,N_{\alpha}=\alpha^{\dagger}\cdot\alpha\,,\cdots$
the number operators associated to a given oscillator. $U$ denotes the
$\mathfrak{u}(1)_r$ Abelian generator. Similarly, pure bosonic or fermionic
bi-oscillators defining $\mathfrak{g}^{\pm}$ correspond to R-charge ladder
operators, translations, and special conformal transformations:
\begin{align}
	\mathfrak{g}^{+}:  &  \qquad K_{i\bar{k}}=a_{i}\cdot b_{\bar{k}}\,,\qquad
	J^{+}=\alpha\cdot\beta\,;\\
	\mathfrak{g}^{-}:  &  \qquad P^{i\bar{k}}=a^{i}\cdot b^{\bar{k}}\,,\qquad
	J^{-}=\beta^{\dagger}\cdot\alpha^{\dagger}\,.
\end{align}
Finally, generators involving both fermionic and bosonic oscillators correspond
to the 8+8 super(conformal) charges:
\begin{align}
Q^{i+}  &  =a^{i}\cdot\alpha\,, & Q^{i-}  &  =a^{i}\cdot\beta^{\dagger}\,, &
\bar{Q}^{\bar{k}+}  &  =b^{\bar{k}}\cdot\beta\,, & \bar{Q}^{\bar{k}-}  &
=b^{\bar{k}}\cdot\alpha^{\dagger}\,,\\
S_{i-}  &  =a_{i}\cdot\alpha^{\dagger}\,, & S_{i+}  &  =a_{i}\cdot\beta\,, &
\bar{S}_{\bar{k}-}  &  =b_{\bar{k}}\cdot\beta^{\dagger}\,, & \bar{S}_{\bar
{k}+}  &  =b_{\bar{k}}\cdot\alpha\,,
\end{align}
The notation is chosen such that $\pm$ indicates the $\mathfrak{su}(2)_{R}$
charge, e.g. $[J^{(3)},Q^{i\pm}]=\pm Q^{i\pm}$, and the normalization of $U$
reproduces the usual conventions. Indeed, using the commutation relations
involving the R-symmetry generators, it is straightforward to check that the
supercharges arrange themselves into the usual
$\mathfrak{su}(2)_{R}\oplus\mathfrak{u}(1)_{r}$ representations.

\paragraph{Commutation relations:} we collect here the commutation relations
between all generators of $\mathfrak{su}(2,2|2)$.

\begin{itemize}
	\item Relations involving the Lorentz generators, $\mathcal{L}^{i}_j$:
		\begin{gather}
			[\mathcal{L}^i_j, \mathcal{L}^k_l] = \delta^k_j\mathcal{L}^i_l - \delta^i_l\mathcal{L}^k_j\,,\qquad
			[\mathcal{L},\bar{\mathcal{L}}]=0=[\mathcal{L},J]\,,\\
			[\mathcal{L}^i_j, K_{k\bar{l}}] = - \delta^i_k K_{j\bar{l}} + \frac{1}{2}\delta^i_j K_{k\bar{l}}\,,\qquad
			[\mathcal{L}^i_j, P^{k\bar{l}}] = \delta^k_j P^{i\bar{l}} - \frac{1}{2}\delta^i_j P^{k\bar{l}}\,,\\
			[\mathcal{L}^i_j, Q^{k\pm}] = +\delta^k_jQ^{i\pm} - \frac{1}{2}\delta^i_j Q^{k\pm}\,,\qquad
			[\mathcal{L}^i_j, S_{k\pm}] = - \delta^i_kS_{j\pm} + \frac{1}{2}\delta^i_j S_{k\pm}\,,\\
			[\mathcal{L}^i_j, \bar{Q}] = 0 = [\mathcal{L}^i_j, \bar{S}]\,.
		\end{gather}
	\item Relations involving the Lorentz generators, $\bar{\mathcal{L}}^{\bar{k}}_{\bar{l}}$:
		\begin{gather}
			[\bar{\mathcal{L}}^{\bar{k}}_{\bar{l}}, \bar{\mathcal{L}}^{\bar{m}}_{\bar{n}}] = \delta^{\bar{m}}_{\bar{l}}\bar{\mathcal{L}}^{\bar{k}}_{\bar{n}} - \delta^{\bar{k}}_{\bar{n}}\bar{\mathcal{L}}^{\bar{m}}_{\bar{l}}\,,\qquad
			[\bar{\mathcal{L}}, \mathcal{L}]=0=[\bar{\mathcal{L}}, J]\,,\\
			[\bar{\mathcal{L}}^{\bar{k}}_{\bar{l}}, K_{i\bar{m}}] = - \delta^{\bar{k}}_{\bar{m}} K_{i\bar{l}} + \frac{1}{2}\delta^{\bar{k}}_{\bar{l}} K_{i\bar{m}}\,,\qquad
			[\bar{\mathcal{L}}^{\bar{k}}_{\bar{l}}, P^{i\bar{m}}] = \delta_{\bar{l}}^{\bar{m}} P^{i\bar{k}} - \frac{1}{2}\delta^{\bar{k}}_{\bar{l}} P^{i\bar{m}}\,,\\
			[\bar{\mathcal{L}}^{\bar{k}}_{\bar{l}}, Q^{i\pm}] = 0 = [\bar{\mathcal{L}}^{\bar{k}}_{\bar{l}}, S_{i\pm}] \,,\\
			[\bar{\mathcal{L}}^{\bar{k}}_{\bar{l}}, \bar{Q}^{\bar{m}\pm}] = \delta^{\bar{m}}_{\bar{l}}\bar{Q}^{\bar{k}\pm} - \frac{1}{2}\delta^{\bar{k}}_{\bar{l}} \bar{Q}^{\bar{m}\pm}\,,\qquad
			[\bar{\mathcal{L}}^{\bar{k}}_{\bar{l}}, \bar{S}_{\bar{m}\pm}] = -\delta^{\bar{k}}_{\bar{m}}\bar{S}_{\bar{l}\pm} + \frac{1}{2}\delta^{\bar{k}}_{\bar{l}} \bar{S}_{\bar{m}\pm}\,.
		\end{gather}
	\item Relations involving special conformal transformations and translations, $K_{i\bar{k}}, P^{i\bar{k}}$:
		\begin{gather}
			[K_{i\bar{k}}, P^{j\bar{l}}] = \delta^{\bar{l}}_{\bar{k}}\mathcal{L}^j_i + \delta^j_i\bar{\mathcal{L}}^{\bar{l}}_{\bar{k}} + \delta^j_i\delta^{\bar{l}}_{\bar{k}}\mathcal{D}\,,\\
			[K_{i\bar{k}}, Q^{j\pm}] = +\delta^j_i \bar{S}_{\bar{k}\pm}\,,\qquad
			[K_{i\bar{k}}, \bar{Q}^{\bar{l}\pm}] = +\delta^{\bar{l}}_{\bar{k}} S_{i\pm}\,,\\
			[K_{i\bar{k}}, S_{j\pm}] = 0 =	[K_{i\bar{k}}, \bar{S}_{\bar{l}\pm}]\,,\qquad
			[P^{i\bar{k}}, Q^{j\pm}] = 0 =	[P^{i\bar{k}}, \bar{Q}^{\bar{l}\pm}]\,,\\
			[P^{i\bar{k}}, S_{j\pm}] = -\delta^i_j \bar{Q}^{\bar{k}\pm}\,,\qquad
			[P^{i\bar{k}}, \bar{S}_{\bar{l}\pm}] = -\delta^{\bar{k}}_{\bar{l}} Q^{i\pm}\,,
		\end{gather}
	\item Relations involving the $\mathfrak{su}(2)_R$ generators, $J^\pm\,,J^{(3)}$:
		\begin{gather}
			[J^+, J^-] = J^{(3)}\,,\\
			[J,\mathcal{L}]=0=[J,\bar{\mathcal{L}}]\,,\qquad
			[J,K]=0=[J,P]\,,\\
			[J^\pm,Q^{i\pm}] = 0 = [J^\pm,\bar{Q}^{\bar{k}\pm}]\,,\qquad
			[J^\pm,S_{i\pm}] = 0 = [J^\pm,\bar{S}_{\bar{k}\pm}]\,,\\
			[J^\pm,Q^{i\mp}] = +Q^{i\pm} \,,\qquad
			[J^\pm,S_{i\mp}] = -S^{i\pm} \,,\\
			[J^\pm,\bar{Q}^{\bar{k}\mp}] = -\bar{Q}^{\bar{k}\pm} \,,\qquad
			[J^\pm,\bar{S}_{\bar{k}\mp}] = +\bar{S}_{\bar{k}\pm} \,,\\
			[J^{(3)}, Q^{i\pm}] = \pm Q^{i\pm}\,,\qquad
			[J^{(3)}, S_{i\pm}] = \pm S_{i\pm}\,,\\
			[J^{(3)}, \bar{Q}^{\bar{k}\pm}] = \pm \bar{Q}^{\bar{k}\pm}\,,\qquad
			[J^{(3)}, \bar{S}_{\bar{k}\pm}] = \pm \bar{S}_{\bar{k}\pm}\,,
		\end{gather}
	\item Anti-commutation relations of the super(conformal) generators:
		\begin{gather}
				\{ Q^{i\pm},S_{j\mp} \} =\mathcal{L}^i_j + \frac{1}{2}\delta^i_j(\mathcal{D}\mp J^{(3)} + \frac{1}{2}U)\,,\qquad
				\{ Q^{i\pm},Q^{j\pm} \} = 0 = \{ S_{i\pm},S_{j\pm} \} \,,\\
				\{\bar{Q}^{\bar{k}\pm},\bar{S}_{\bar{l}\mp}\} = \bar{\mathcal{L}}^{\bar{k}}_{\bar{l}} + \frac{1}{2}\delta^{\bar{k}}_{\bar{l}}(\mathcal{D}\mp J^{(3)}-\frac{1}{2}U)\,,\qquad
				\{ \bar{Q}^{\bar{k}\pm},\bar{Q}^{\bar{l}\pm} \} = 0 =  \{ \bar{S}_{\bar{k}\pm},\bar{S}_{\bar{l}\pm} \}\,,\\
				\{Q^{i\pm},\bar{Q}^{\bar{k}\mp} \} = P^{i\bar{k}}\,,\qquad
				\{S_{i\mp},\bar{S}_{\bar{k}\pm} \} = K_{i\bar{k}}\,,\\
				\{Q^{i\pm},S_{j\pm} \} = -\delta^i_j J^{\pm}\,,\qquad
				\{\bar{Q}^{\bar{k}\pm},\bar{S}_{\bar{l}\pm} \} = +\delta^{\bar{k}}_{\bar{l}} J^{\pm}\,,\qquad
		\end{gather}
	\item Relations involving the generators $\mathcal{D}$ and $U$:
		\begin{gather}
			[\mathcal{D}, \mathcal{L}]=0=[\mathcal{D},\bar{\mathcal{L}}]\,,\qquad [\mathcal{D}, K] = -K\,,\qquad [\mathcal{D}, P] = P\,,\\
			[U,\mathcal{L}] = 0 = [U,\bar{\mathcal{L}}]\,,\qquad
   			[U,P] = 0 = [U,K]\,,\\
			[U, Q] = -Q\,,\qquad [U, S]=+S\,,\\
			[U, \bar{Q}] = +\bar{Q}\,,\qquad [U, \bar{S}]= -\bar{S}\,,
		\end{gather}
\end{itemize}

\begin{table}[ptb]
\centering
\begin{tabular}[c]{c|c|c|}
	field & state & $[\Delta;\ell,\bar{\ell};j,u]$\\\hline
	$X$ $(*)$ & $\left|  0\right>  $ & $[1; 0,0; 1,0]$\\
	$Y^{\dagger}$ & $\beta^{\dagger}\alpha^{\dagger}\left|  0\right>  $ & $[1;0,0;-1,0]$\\
	$\psi_{X}^{i}$ & $a^{i}\beta^{\dagger}\left|  0\right>  $ & $[\frac{1}{2};\pm1,0; 0,-1]$\\
	$\bar{\psi}_{Y}^{\bar{k}}$ & $b^{\bar{k}}\alpha^{\dagger}\left|  0\right>$ & $[\frac{1}{2};0,\pm1; 0,+1]$\\\hline
	$\varphi$ $(*)$ & $\alpha^{\dagger}\left|  0\right>  $ & $[1; 0,0; 0, 2]$\\
	$\chi^{i}_{+}$ & $a^{i}\left|  0\right>  $ & $[\frac{1}{2}; \pm1,0; 1, 1]$\\
	$\chi^{i}_{-}$ & $a^{i}\beta^{\dagger}\alpha^{\dagger}\left|  0\right>$ & $[\frac{1}{2}; \pm1,0; -1, 1]$\\
	$F^{ij}$ & $a^{i}a^{j}\beta^{\dagger}\left|  0\right>  $ & $[2; 2,0; 0,	0]$\\\hline
	$\bar{\varphi}$ $(*)$ & $\alpha^{\dagger}\left|  0\right>  $ & $[1; 0,0; 0,	-2]$\\
	$\bar{\chi}^{\bar{k}}_{+}$ & $b^{\bar{k}}\left|  0\right>  $ & $[\frac{1}{2}; 0, \pm1; 1, -1]$\\
	$\bar{\chi}^{\bar{k}}_{-}$ & $b^{\bar{k}}\beta^{\dagger}\alpha^{\dagger}\left|0\right> $ & $[\frac{1}{2}; \pm1,0; -1, -1]$\\
	$\bar{F}^{\bar{k}\bar{l}}$ & $b^{\bar{k}}b^{\bar{l}}\alpha^{\dagger}\left|0\right>  $ & $[2; 0,2; 0, 0]$
\end{tabular}
\caption{Conformal primaries of spin smaller than two, their oscillator representations,
and their quantum numbers, grouped according to their $\mathcal{N}=2$
superconformal multiplets. The superconformal primary of the free
hypermultiplet, $\widehat{\mathcal{B}}_{1}$, the $\mathcal{N}=2$ vector
multiplets, $\mathcal{E}_{2}$, $\bar{\mathcal{E}}_{-2}$, are signaled by an asterisk
$(*)$. For vector fields and spinors, we have only indicated
the quantum numbers of their Lorentz highest weight states.}
\label{tab:N=2fields}
\end{table}

\paragraph{Representations of Superconformal Multiplets:} To make contact with
the familiar fields of $\mathcal{N}=2$ theories in the oscillator
representation, let us first define the quantum numbers of the 4D
superconformal group, corresponding to the decomposition of the superconformal
algebra in terms of its bosonic subalgebra:
\begin{equation}\label{4d-bosonic-decomposition}
		\mathfrak{su}(2,2|2)\longrightarrow \mathfrak{so}(2)\oplus\mathfrak{su}(2)\oplus\mathfrak{su}(2)\oplus\mathfrak{su}(2)_R\oplus\mathfrak{u}(1)_r\,:\qquad [\Delta;\ell,\bar{\ell}; j, r]\,.
\end{equation}
To that end, we define the following combinations of the Lorentz generators:
\begin{gather}
	\mathcal{L}^{+}=\mathcal{L}_{2}^{1}\,,\qquad
	\mathcal{L}^{-}=\mathcal{L}_{1}^{2}\,,\qquad
	\mathcal{L}^{(3)}=\mathcal{L}_{1}^{1} - \mathcal{L}_{2}^{2} = a^{1}\cdot a_{1}-a^{2}\cdot a_{2}\,,\\
	[\mathcal{L}^{+},\mathcal{L}^{-}]=\mathcal{L}^{(3)}\,,\qquad
	[\mathcal{L}^{(3)},\mathcal{L}^{\pm}]=\pm2\mathcal{L}^{\pm}\,,\qquad
\end{gather}
and similarly for $\bar{\mathcal{L}}_{\bar{l}}^{\bar{k}}$. These three
generators clearly satisfy both the $\mathfrak{su}(2)$ commutation relations
and the Serre--Chevalley relations, see equation \eqref{Serre--Chevalley}, and
therefore form a useful basis of
$\mathfrak{so}(4)=\mathfrak{su}(2)\oplus\mathfrak{su}(2)$. The quantum numbers
associated to a state $\left\vert \Omega\right\rangle $ are then given by:
\begin{equation}\label{4d-quantum-numbers-1}
	\mathcal{L}^{(3)}\left| \Omega\right> =\ell\,\left|\Omega\right> \,,\qquad
	\bar{\mathcal{L}}^{(3)}\left|\Omega\right> = \bar{\ell}\,\left| \Omega\right> \,,\qquad
	\mathcal{D}\left|\Omega\right> =\Delta\,\left|\Omega\right> \,,
\end{equation}
while the charges of the $\mathfrak{su}(2)_{R}\oplus\mathfrak{u}(1)_{r}$
R-symmetry are obtained via the Cartan elements:\footnote{The Cartan elements
for all $\mathfrak{su}(2)$ also follow the mathematics convention and have
integer eigenvalues. This make many computations, such as finding the Cartan
matrices, easier.}
\begin{equation}\label{4d-quantum-numbers-2}
	J^{(3)}\left|\Omega\right> = j\,\left| \Omega\right>\,,\qquad
	U\left|\Omega\right> = r \,\left| \Omega\right>\,.
\end{equation}
Any operator of the 4D SCFT can be represented in terms of oscillators and
labeled by the quantum numbers. As in 6D, the vacuum itself is associated with
the scalar highest weight state of the 4D $\mathcal{N}=2$ hypermultiplet:
\begin{equation}
		X\longleftrightarrow\left\vert 0\right\rangle: \qquad [\Delta;\ell,\bar{\ell};j,r] = [1;0,0;1,0]\,.
\end{equation}
As expected, it is annihilated by all superconformal charges $S_{i\pm},
\bar{S}_{\bar{k}\pm}$, and is therefore a superconformal primary. The
$\mathcal{N}=2$ hypermultiplet can be further split in two $\mathcal{N}=1$
chiral/antichiral multiplet $(X,\psi_X)\oplus (Y^\dagger, \bar{\psi}_Y)$,
containing a Weyl fermion of each chirality, which can be obtained by acting on
the oscillator vacuum with the supercharges and $J^-$:
\begin{equation}
	Y^{\dagger}\leftrightarrow J^{-}\left|  0\right> \,,\qquad
	\psi_{X}^{i}\leftrightarrow Q^{i-}\left|  0\right>\,,\qquad
	\bar{\psi}_{Y}^{\bar{k}}\leftrightarrow\bar{Q}^{\bar{k}-}\left|0\right>\,,
\end{equation}
Derivatives of these fields are generated with $P^{i\bar{k}}$:
\begin{equation}
		D^{i\bar{k}}X\leftrightarrow P^{i\bar{k}}\left|  0\right>\,,\quad
		D^{j\bar{l}}D^{i\bar{k}}X\leftrightarrow P^{j\bar{l}}P^{i\bar{k}}\left|0\right>\,,\quad\cdots\quad D^{i\bar{k}}Y^{\dagger}\leftrightarrow P^{i\bar{k}}J^{-}\left|  0\right>  \,,\quad\cdots \,,
\end{equation}
to form, as expected, a complete $\widehat{\mathcal{B}}_{1}$ superconformal
multiplet in the language of reference \cite{Dolan:2002zh}.\footnote{The
		nomenclature of \cite{Dolan:2002zh} is most often used in the
		$\mathcal{N}=2$ SCFT literature.  In the nomenclature of reference
		\cite{Cordova:2016emh}, which is more appropriate to diverse
		dimensions, and which we adopt when discussing six-dimensional
		spin chains, the free $\mathcal{N}=2$ hypermultiplet is defined as a
$B_{1}\bar{B}_{1}[0;0]^{(1;0)}_{1}$ superconformal multiplet.}

Note that the oscillator construction of free fields is independent of the type
of system considered, and we can also consider other types of $\mathcal{N}=2$
multiplets in the free field limit, although they will be strongly coupled in
the kind of analyses done in this work. For completeness, we give them in Table
\ref{tab:N=2fields}.

\begin{figure}[t]\label{}
        \centering
	\begin{subfigure}[t]{.4\linewidth}
		\centering
		{\vspace{0.4cm}\resizebox{0.65\textwidth}{!}{\includegraphics{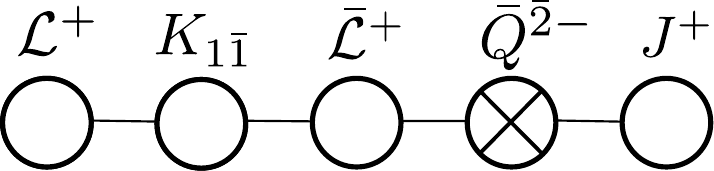}}}
		\caption{Distinguished diagram}
		\label{fig:dynkin_4d_9}
	\end{subfigure}
	\begin{subfigure}[t]{.4\linewidth}
		\centering
		{\vspace{0.4cm}\resizebox{0.65\textwidth}{!}{\includegraphics{figures/dynkin_4d_8.pdf}}}
		\caption{The ``Beauty''}
		\label{fig:dynkin_4d_8}
	\end{subfigure}
	\begin{subfigure}[t]{.4\linewidth}
		\centering
		{\vspace{0.65cm}\resizebox{0.65\textwidth}{!}{\includegraphics{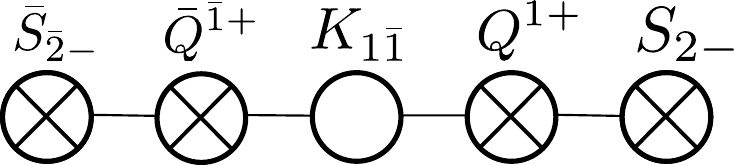}}}
		\caption{}
		\label{fig:dynkin_4d_1}
	\end{subfigure}
	\begin{subfigure}[t]{.4\linewidth}
		\centering
		{\vspace{0.65cm}\resizebox{0.65\textwidth}{!}{\includegraphics{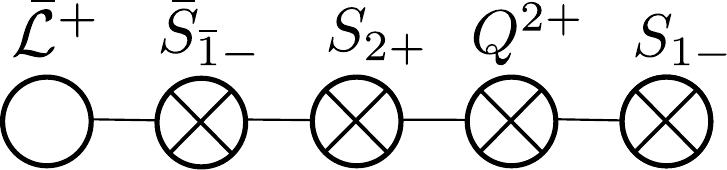}}}
		\caption{}
		\label{fig:dynkin_4d_2}
	\end{subfigure}
	\begin{subfigure}[t]{.4\linewidth}
		\centering
		{\vspace{0.5cm}\resizebox{0.65\textwidth}{!}{\includegraphics{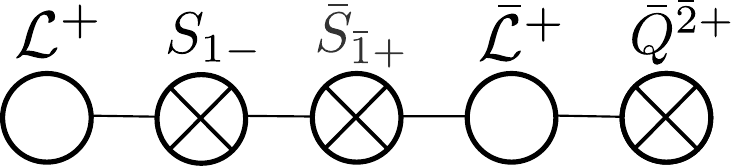}}}
		\caption{}
		\label{fig:dynkin_4d_3}
	\end{subfigure}
	\begin{subfigure}[t]{.4\linewidth}
		\centering
		{\vspace{0.5cm}\resizebox{0.65\textwidth}{!}{\includegraphics{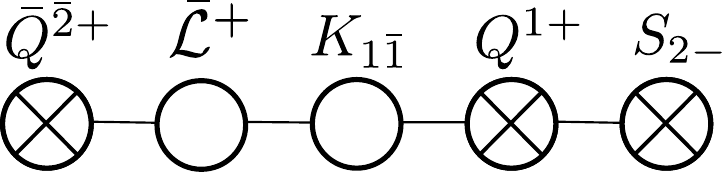}}}
		\caption{}
		\label{fig:dynkin_4d_5}
	\end{subfigure}
	\begin{subfigure}[t]{.4\linewidth}
		\centering
		{\vspace{0.5cm}\resizebox{0.65\textwidth}{!}{\includegraphics{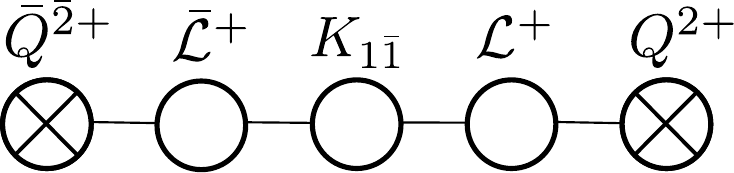}}}
		\caption{}
		\label{fig:dynkin_4d_6}
	\end{subfigure}
	\begin{subfigure}[t]{.4\linewidth}
		\centering
		{\vspace{0.5cm}\resizebox{0.65\textwidth}{!}{\includegraphics{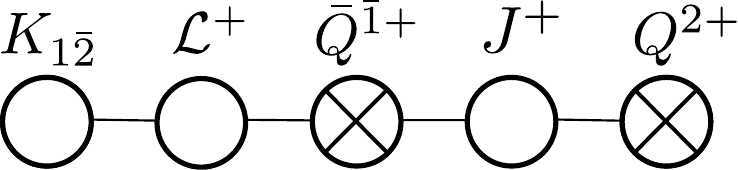}}}
		\caption{}
		\label{fig:dynkin_4d_7}
	\end{subfigure}
	\caption{
		\label{fig:4DDynkin}
		Topologically distinct Dynkin diagrams of the 4D
		superconformal algebra, $\mathfrak{su}(2,2|2)$, with each node labeled
		by the associated generator, $E^+_I$. Fermionic nodes are denoted by a
		cross. The representative is chosen such that $X$ is the highest weight
		state of its representation, except for the distinguished diagram,
		where we did not find such a representative.
	}
\end{figure}

\paragraph{Root Systems and Dynkin Diagrams:} With the same method we used in
6D, we scanned for all possible root systems to find the different Dynkin
diagrams of $\mathfrak{su}(2,2|2)$, which results in the eight Dynkin diagrams
collected in Figure \ref{fig:4DDynkin}. The distinguished diagram is given in
Figure \ref{fig:dynkin_4d_9}, while the ``Beauty'', making the maximal compact
subalgebra,
$\mathfrak{su}(2|1)\oplus\mathfrak{su}(2|1)\oplus\mathfrak{u}(1)_r$, manifest
is shown in Figure \ref{fig:dynkin_4d_8}. Note that in some cases, we did not
find representatives of the Dynkin diagram where $X$ is a highest weight state
with respect to the associated root system. This happens, for instance, with the
distinguished diagram. In those cases, the half-BPS states are interpreted as a
highly-excited operator of the super-spin chain rather than its ground state.
This also happens for $\mathcal{N}=4$ super Yang--Mills, which has led to the
nomenclature of Beauty and the Beast for the root systems
\cite{Beisert:2003yb}. Despite this difference, the Bethe ans\"atze are nevertheless
equivalent, since there exists a super-Weyl transformation relating them.

\bibliographystyle{utphys}
\bibliography{6Dsuperchain}

\end{document}